\documentclass{amsart}
\usepackage{ae,aecompl}\usepackage{ifthen}
\usepackage[a4paper,twoside]{geometry}
\usepackage{amsmath, amsthm}
\usepackage{graphicx}
\usepackage{amssymb}
\usepackage{url}
\newboolean{dum}
\setboolean{dum}{false}
\newcommand{\comment}[1]{\ifthenelse{\boolean{dum}}{
{\par\noindent\Huge\ding{46}} \fbox{\parbox{10cm}{#1}}\par}{}}

\newtheorem{lemma}[subsection]{Lemma}

\newtheorem{prop}[subsection]{Proposition}
\newtheorem{cor}[subsection]{Corollary}

\numberwithin{equation}{section}

\begin{document}
\title{Eigenvalue separation in some random matrix models}
\author{K.E.~Bassler}
\address{
Department of Physics 
and 
Texas Center for Superconductivity,
University of Houston,
Houston, Texas 77204, USA}
\author{P.J.~Forrester}
\address{Department of Mathematics and Statistics,
The University of Melbourne,
Victoria 3010, Australia}
\author{N.E.~Frankel}
\address{School of Physics,
The University of Melbourne,
Victoria 3010, Australia}
\begin{abstract}{The eigenvalue density for members of the Gaussian orthogonal
and unitary ensembles follows the Wigner semi-circle law. If the Gaussian
entries are all shifted by a constant amount $c/(2N)^{1/2}$, where $N$ is the size
of the matrix, in the large $N$ limit a single eigenvalue will separate
from the support of the Wigner semi-circle provided $c > 1$. In this study,
using an asymptotic analysis of the secular equation for the eigenvalue condition, 
we 
compare this effect
to analogous effects occurring in general variance 
Wishart matrices and matrices from the shifted mean chiral ensemble.
We undertake an analogous comparative study of eigenvalue
separation properties when 
the size of the matrices are fixed and $c \to \infty$, and higher
rank analogues of this setting. This is done using
exact expressions for eigenvalue probability densities in terms of generalized
hypergeometric functions, and using the interpretation of the latter
as a Green function in
the Dyson Brownian motion model.
For the shifted mean Gaussian unitary ensemble and its analogues an alternative approach
is to use exact expressions for the correlation functions in terms
of classical orthogonal polynomials and associated multiple
generalizations. By using these exact expressions to compute and plot
the eigenvalue density, illustrations of the various eigenvalue
separation effects are obtained.
}
\end{abstract}
\maketitle

\section{Introduction}
\subsection{The aims of the paper}
Our interest is in the shifted mean Gaussian ensemble, the general variance
Laguerre ensemble, and the shifted mean chiral ensembles from random matrix
theory. The shifted mean Gaussian ensemble consists of real symmetric
$(\beta = 1)$ or complex Hermitian $(\beta = 2)$ matrices $H$ with joint
distribution of the elements proportional to
\begin{equation}\label{1.1}
\exp \Big ( - {\beta \over 2} {\rm Tr} ( H - H^{(0)})^2 \Big )
\end{equation}
where $H^{(0)}$ is a fixed matrix of the same type as $H$. It follows
immediately from (\ref{1.1}) that $H = G + H^{(0)}$ where $G$ is a member of
the (zero mean) Gaussian orthogonal ensemble ($\beta = 1$) or Gaussian unitary
ensemble $(\beta = 2$). In the early stages
of the development of random matrix theory in nuclear physics, Porter studied
the real case of (\ref{1.1}), with all entries of $ H^{(0)}$ equal to
a constant $\mu$, through computer experiments
of the eigenvalue distribution. It was found that typically all but one of the
eigenvalues followed the prediction of the Wigner semi-circle law (see e.g.~\cite{Fo02}),
and so to leading order were supported on $[-\sqrt{2N}, \sqrt{2N}]$.
The remaining eigenvalue---the largest---separated from the semi-circle. We owe our
knowledge of this to the work of Lang \cite{La64}, who followed up on this
empirical finding with a theoretical study predicting that the largest eigenvalue
would occur near $\mu N$.

Jones et al.~\cite{JKT78}, by way of their analysis of a spherical model of spin glasses
\cite{KTJ78}, made a quantitative study of this effect in the large $N$ limit.
They showed (see Section
\ref{s2} below) that with $H^{(0)}$ having all elements equal to $c/\sqrt{2N}$,
the separation of the largest eigenvalue only occurs for $c > 1$, thus exhibiting
a phase transition in $c$.

Many year later \cite{BBP05}, an analogous effect has been exhibited in the
general variance Laguerre (also known as Wishart) ensemble. This ensemble consists
of matrices $X^\dagger X$ (Wishart matrices) where the joint distribution on the
elements of the $n \times m$ $(n \ge m)$ matrix $X$ is proportional to
\begin{equation}\label{1.2}
\exp \Big ( - {\beta \over 2} {\rm Tr} (X^\dagger X \Sigma^{-1}) \Big )
\end{equation}
where $\Sigma$ is the $m \times m$ covariance matrix, and as such is required to
have all eigenvalues positive. The elements of $X$ are real for $\beta = 1$, and
complex for $\beta = 2$. Writing $Y = X \Sigma^{-1/2}$ we see $Y^\dagger Y$
is a Wishart matrix with $\Sigma = {\bf 1}_m$, and furthermore
\begin{equation}\label{1.2a}
X^\dagger X = \Sigma^{1/2} Y^\dagger Y \Sigma^{1/2} \sim Y^\dagger Y \Sigma
\end{equation}
where the tilde denotes the matrices are similar and so have the same
eigenvalues. With
\begin{equation}\label{3.0}
\Sigma = {\rm diag}(b, (1)^{m-1})
\end{equation}
(here the notation $(1)^{m-1}$ denotes 1 repeated $m-1$ times) it was shown that the largest
eigenvalue separates from the leading support on $(0,4m)$ provided $bm > 2$.

It is the objective of this paper to exhibit the mathematical similarities between these two
eigenvalue separation phenomena. Moreover, we add to this a third example, namely the shifted
mean chiral matrices
\begin{equation}\label{3.1}
\left [ \begin{array}{cc} 0_{m \times m} & X + X^{(0)} \\
X^\dagger + (X^{(0)})^\dagger & 0_{n \times n} \end{array} \right ]
\end{equation}
where $X$ is an $n \times m$ Gaussian random matrix with distribution given by (\ref{1.2})
in the case $\Sigma = {\bf 1}_m$. The matrix $X^{(0)}$ is a constant matrix of the
same type as $X$. With all elements of $X^{(0)}$ equal to $c/(2\sqrt{m})$, a separation of the
largest eigenvalue in modulus (the eigenvalues of (\ref{3.1}) occur in $\pm$ pairs) only
occurs for $c > 1$.

\subsection{Related works}
In the mathematics literature the problem of the statistical properties of
the largest eigenvalue of (\ref{1.1}) in the case that all entries of
$H^{(0)}$ are equal to $\mu$ was first studied by F\"uredi and Komlos \cite{FK81}.
In the more general setting of real Wigner matrices (independent entries i.i.d.~with
mean $\mu$ and variance $\sigma^2$) the distribution of the largest eigenvalue
was identified as a Gaussian, so generalizing the result of Lang.
Only in recent years did the associated phase transition problem, already
known to Jones et al.~\cite{JKT78}, receive attention in the mathematical
literature. In the case of the GUE, this was due to Pech\'e \cite{Pe06},
while a rigorous study of the GOE case can be found in the work of Maida \cite{Ma07}.

The paper \cite{BBP05} by Ben Arous, Baik and Pech\'e proved the phase transition property
relating to (\ref{1.2}) in the complex case with $\Sigma$ given by 
$\Sigma = {\rm diag} ((b)^r, (1)^{m-r})$  ((\ref{3.0}) corresponds to $r=1$).
Subsequent studies by Baik and Silverstein \cite{BS06}, Paul \cite{Pa07} and
Bai and Yao \cite{BY08} considered the real case. Significant for the present study is the
result of \cite{BY08}, giving that for (\ref{1.2}) with $\Sigma$ given as above, 
the separated eigenvalues have the law of the $r \times r$ GUE. The case $\beta = 4$---not
considered here---corresponding to self dual quaternion real matrices,
is studied in the recent work of Wang \cite{Wa07}.

The eigenvalue probability density function (\ref{1.1}) is closely related to the Dyson Brownian motion model
\cite{Dy62b} in random matrix theory (see Section 3 below). It is also referred to
as a Gaussian ensemble with a source. In this context the case of $H^{(0)}$ having
a finite rank has been studied by a number of authors \cite{BK04a,IS05,BL08,ADV07}.
However our use of this differs in that we will keep $N$ fixed, and exhibit phase
separation as a function of the perturbing parameter.

\subsection{Plan}
We begin in Section 2 by showing how in each case the criteria for separation can be
deduced from the eigenvalue equation implied by regarding the parameters $b,c,$ as the
proportionalities in rank 1 (or in the case of (\ref{3.1}), rank 2) perturbations.
In Section 3 we discuss analytic features of the eigenvalue separation effect in the case that
the matrices $H^{(0)}$, $\Sigma$ and $X$ are of rank $r$. Working with matrices of fixed size,
we make use of a generalized hypergeometric function form of the joint
eigenvalue probability density function.

In Section 4, in the case $\beta = 2$ (complex) case, we draw attention
to known explicit forms for the $k$-point correlation functions, and show
how these are well suited to exhibiting eigenvalue separation both analytically
and numerically. The content of Section 5 is to  exhibit an analogue of eigenvalue separation
for the general variance Laguerre ensemble, variance matrices of the form
(\ref{3.0}), and its generalization $\Sigma = {\rm diag}((b)^r, (1)^{m-r})$ in the limit
$b \to 0$. An applied setting is identified in Section 6 as a motivation for future
work.

\section{Secular equations}\label{s2}
Consider first the shifted mean Gaussian ensemble (\ref{1.1}), and for
definiteness suppose $\beta = 1$ ($N \times N$ real symmetric matrices).
The simplest case is when all elements of $H_0$ are constant, equal to $\mu$ say.
Then
\begin{equation}\label{HAx}
 H = G + \mu \vec{x} \vec{x}^T
\end{equation}
where $\vec{x}$ is a column vector with all entries equal to 1, and $G$ is a member of
the (zero mean) Gaussian orthogonal ensemble.
Diagonalizing $G$,
$G = O  L  O^T$, $ L = {\rm diag} \, (a_1,\dots,a_N)$, and using the
fact that $ O^T \vec{x} =: \vec{y}$ is distributed as a standard Gaussian vector shows
that from the viewpoint of the eigenvalues, the right hand side of (\ref{HAx}) can be replaced by
$ L + \mu \vec{y} \vec{y}^T$. We seek the eigenvalues of this matrix \cite{An96}.

\begin{lemma}\label{p1.18sm}
The eigenvalues of the matrix
$$
\tilde{H} := {\rm diag} (a_1,\dots,a_N) + \mu \vec{y} \vec{y}^T
$$
are given by the zeros of the solution of the secular equation
(for the use of this terminology, see \cite{An96})
\begin{equation}\label{1.dAm}
0 = 1 - \mu \sum_{i=1}^N {y_i^2 \over \lambda - a_i}.
\end{equation}
Assuming the ordering $a_1> \cdots > a_N$, and that $\mu >0$, a corollary
of this is that the eigenvalues satisfy the interlacing
\begin{equation}\label{1.intl}
\lambda_1 > a_1 > \lambda_2 > a_2 > \cdots > \lambda_N > a_N.
\end{equation}
\end{lemma}

\noindent{\it Proof.} \quad
With $A = {\rm diag}(a_1,\dots,a_N)$ we have
\begin{equation}\label{1.dAH}
\det( {\bf 1}_N \lambda - \tilde{H}) =
\det( {\bf 1}_N \lambda - {A})
\det(  {\bf 1}_N - \mu \vec{y} \vec{y}^T
({\bf 1}_N \lambda - {A})^{-1} ).
\end{equation}
The matrix product in the second determinant has rank 1 and so
\begin{equation}\label{1.dAH1}
\det(  {\bf 1}_N - \mu \vec{y} \vec{y}^T
({\bf 1}_N  \lambda - {A})^{-1} ) = 1- {\rm Tr} (  \mu \vec{y} \vec{y}^T
({\bf 1}_N \lambda - { A})^{-1} ) =  1 - \mu \sum_{i=1}^N {y_i^2 \over \lambda - a_i}.
\end{equation}
The characteristic polynomial (\ref{1.dAH}) vanishes at the zeros of this determinant,
but not at the zeros of $\det({\bf 1}_N \lambda - {A})$ due to the cancellation with the
poles in
(\ref{1.dAH1}). Thus the condition for the eigenvalues is the secular equation
(\ref{1.dAm}). The interlacing condition can
be seen by sketching a graph, taking into consideration the requirement  $\mu y_i^2 > 0$.
\hfill $\square$

We are interested in the 
position of the largest eigenvalue of
$\tilde{H}$ or equivalently $H$
as a function of $\mu$ and $N$, which unlike the other eigenvalues is not
trapped by the eigenvalues of $A$. Following \cite{JKT78}, to do this requires
using the Wigner semi-circle law for the eigenvalue density of Gaussian orthogonal
ensemble matrices,
\begin{equation}\label{ws}
\rho_{(1)}^{\rm W}(x) \sim \left \{
\begin{array}{ll} (2N/\pi J)(1 - x^2/J^2), & |x| < J \\
0, & |x| > J \end{array} \right. 
\end{equation}
where $J = (2N)^{1/2}$. Thus in (\ref{1.dAm}) we begin by averaging over the components
of the eigenvectors and thus replacing each $y_i^2$ by its
mean value unity. Because the density of $\{a_i\}$ is given by $\rho_{(1)}^{\rm W}(x)$,
for large $N$ (\ref{1.dAm}) then assumes the form of an integral equation,
\begin{equation}\label{3.dyb}
0 = 1 - \mu \int_{-J}^J {\rho_{(1)}^{\rm W}(x) \over \lambda - x} \, dx.
\end{equation}
We seek a solution of this equation for $\lambda > J$. 

Now, for $\lambda > J$,
\begin{equation}\label{3.dyb1}
\int_{-J}^J {\rho_{(1)}^{\rm W}(x) \over \lambda - x} \, dx = {1 \over \lambda}
\sum_{k=0}^\infty {1 \over \lambda^{2k} }
\int_{-J}^J x^{2k} \rho_{(1)}^{\rm W}(x) \, dx.
\end{equation}
But it is well known (see \cite{Fo02}), and can readily be checked directly, that
\begin{equation}\label{3.dyb2}
\int_{-J}^J x^{2k} \rho_{(1)}^{\rm W}(x) \, dx = {2N \over J} (J/2)^{2k+1} c_k
\end{equation}
where $c_k$ denotes the $k$-th Catalan number. Substituting (\ref{3.dyb2}) in (\ref{3.dyb1}),
and using the generating function for the Catalan numbers
$$
\sum_{k=0}^\infty c_k t^k = {1 \over 2t} \Big ( 1 - (1 - 4t)^{1/2} \Big ),
$$
we deduce the integral evaluation
\begin{equation}\label{ie}
\int_{-J}^J {\rho_{(1)}^{\rm W}(x) \over \lambda - x} \, dx = {2N \lambda \over J^2}
\Big ( 1 - (1 - J^2/\lambda^2)^{1/2} \Big ), \qquad |\lambda| > J.
\end{equation}
Substituting this in (\ref{3.dyb}) shows a solution with $\lambda > J$ is only
possible for $\mu > J/(2N)$, and furthermore allows the location of the separated
eigenvalue to be specified.

\begin{prop}\cite{JKT78} \label{p.jk} Consider shifted mean Gaussian orthogonal ensemble matrices
as specified by (\ref{HAx}), and write $\mu = c/J$, $J = (2N)^{1/2}$. For large $N$,
and with $c > 1$, a single eigenvalue separates from the eigenvalue support in the
case $\mu = 0$, $(-J,J)$, and is located at
\begin{equation}\label{cc}
\lambda = {J \over 2} \Big ( c + {1 \over c} \Big ).
\end{equation}
\end{prop}
 
The case of the shifted mean Gaussian ensemble (\ref{1.1}) with $\beta = 2$ ($N \times N$ complex
Hermitian matrices) and all elements of $H_0$ constant, is very similar. With $G$ a
(zero mean) member of the Gaussian unitary ensemble, the shifted mean matrices have the same
eigenvalues distribution as the random rank 1 perturbation
\begin{equation}\label{Gx}
{\rm diag} \, G + \mu \vec{x} \vec{x}^\dagger
\end{equation}
where $\vec{x}$ is an $n$-component standard complex Gaussian vector. Thus with $y_i^2$ replaced by
$|y_i|^2$ the secular equation (\ref{1.dAm}) again specifies the eigenvalues. Since the mean of
$|y_i|^2$ is unity, and the eigenvalue density of Gaussian unity ensemble matrices is again given
by the Wigner semi-circle law (\ref{ws}), all details of the working leading to Proposition
\ref{p.jk} remain valid, and so the proposition itself remains valid for shifted mean unitary
ensemble matrices.

We turn our attention next to correlated Wishart matrices (\ref{1.2a}), and
consider for definiteness the complex case $\beta = 2$. To begin we make use of
the general fact that the non-zero eigenvalues of the matrix products $AB$ and $BA$ are equal.
This tells us that the non-zero eigenvalues of $Y^\dagger Y \Sigma$ are the same as those
for $Y \Sigma Y^\dagger$. With $\Sigma$ given by (\ref{3.0}),
\begin{equation}\label{fns}
Y \Sigma Y^\dagger = \tilde{Y} \tilde{Y}^\dagger + b \vec{y} \vec{y}^\dagger
\end{equation}
where $\vec{y}$ is an $n$-component standard complex Gaussian vector and $\tilde{Y}$ is the
$(n-1) \times m$ matrix constructed from $Y$ by deleting the first column. For
$n > m$ the eigenvalues
of this matrix have the same distribution as
\begin{equation}\label{fn}
{\rm diag} ((0)^{n-1-m}, a_1,\dots, a_m) + b \vec{y} \vec{y}^\dagger
\end{equation}
where $\{(0)^{n-1-m}, a_1,\dots, a_m\}$ are the eigenvalues of $\tilde{Y} \tilde{Y}^\dagger$,
and thus $\{a_1,\dots,a_m\}$ are the eigenvalues of $ \tilde{Y}^\dagger \tilde{Y}$.
 
We recognize (\ref{fn}) as structurally identical to (\ref{Gx}). Hence the condition
for the eigenvalues is given by (\ref{1.dAm}) with $y_i^2$ replaced by $|y_i^2|$,
$\mu$ replaced by $b$,
$N$ replaced by $n-1$, $a_i = 0$ $(i=1,\dots,n-1-m)$, $a_i \mapsto a_{i-(n-1-m)}$
$(i=n-m,\dots, n-1)$. In the limit that $m \to \infty$ with $n-m$ fixed, the
density of eigenvalues for complex Wishart matrices follows the
Mar\u{c}enko-Pastur law (see e.g.~\cite{Fo02})
\begin{equation}\label{fn1}
\lim_{m \to \infty} \rho_{(1)}^{\rm MP}(ym) = \left \{
\begin{array}{ll} (1/\pi \sqrt{y}) (1 - y/4)^{1/2}, & 0 < y < 4 \\
0, & {\rm otherwise}
\end{array} \right.
\end{equation}
Hence, analogous to (\ref{ie}), after averaging and for large $m$ with $n-m$ fixed the
secular equation becomes the integral equation
\begin{equation}\label{ie1}
1 = b m \int_0^{4m} {\rho_{(1)}^{\rm MP}(y) \over \lambda - y} \, dy.
\end{equation}
Substituting (\ref{fn1}), and changing variables $y \mapsto m x^2$ reduces  this
to read
\begin{equation}\label{ie2}
1 = {2bm \over \pi} \int_0^2 {(1 - x^2/4)^{1/2} \over \lambda/m - x^2} \, dx.
\end{equation}
Noting that $(m/\pi) (1-x^2/4)^{1/2} = \rho_{(1)}^{\rm W}(x)$ with
$J=2$, $N=m$ in the latter, following the strategy which lead to (\ref{ie}) 
the integral can be evaluated to give
\begin{equation}\label{2.18}
1 = {bm \over 2} \Big ( 1 - (1 - 4m/\lambda)^{1/2} \Big ).
\end{equation}
This equation only permits a solution for $bm > 2$, which therefore is the condition
for eigenvalue separation. Moreover, the position of the separated eigenvalue can be
read off according to the following result.

\begin{prop}\cite{BBP05} \label{pbib}
Consider the rank 1 perturbation (\ref{fns}) and suppose $b \mapsto b/m$. In the limit
$m \to \infty$ with $n-m$ fixed, provided $b > 2$ a single eigenvalue separates
from the eigenvalue support $(0,4m)$ in the case $b=0$,  and occurs at
$$
\lambda = m {b^2 \over b - 1}.
$$
\end{prop} 

The same result holds for real Wishart matrices. In this case, in (\ref{fns}) we must replace
$\vec{y} \vec{y}^\dagger$ by $\vec{y} \vec{y}^T$ where $\vec{y}$ is a standard real Gaussian
vector, and we must replace $\tilde{Y} \tilde{Y}^\dagger$ by $\tilde{Y} \tilde{Y}^T$ where
$\tilde{Y} \tilde{Y}^T$ is a real Gaussian matrix. Only the mean value of
$y_i^2$, and the density of eigenvalues of $\tilde{Y} \tilde{Y}^T$,
are relevant to the subsequent analysis and as these are unchanged relative to the complex
case the conclusion is the same.

Also of interest in relation to (\ref{fns}) is the limit $m \to \infty$ with $n/m = \gamma > 1$.
This is distinct to the case $n-m$ fixed, because the eigenvalue density is no longer given
by (\ref{fn1}), but rather the more general Mar\u{c}enko-Pastur law (see e.g.~\cite{Fo02})
\begin{equation}\label{MP}
\lim_{m \to \infty} \rho_{(1)}^{\rm MP_\gamma}(x\gamma m) = \Big ( 1 - {1 \over \gamma} \Big ) \delta(x) +
{\sqrt{(x-c)(d-x)} \over 2 \pi \gamma x} \chi_{c < x < d}
\end{equation}
where
$$
c = (1 - \sqrt{\gamma})^2, \qquad d = (1 + \sqrt{\gamma})^2.
$$
Note that the delta function term in (\ref{MP}) is in keeping with the fraction of zero
eigenvalues of $\tilde{Y} \tilde{Y}^\dagger$ equalling $(1 - 1/\gamma)$ 
(recall (\ref{fn})).

In this case the averaged large $m$ limit of the secular equation becomes the integral equation
\begin{equation}\label{s1.1}
1 = {b n (1 - 1/\gamma) \over \lambda} +
{b n \over \gamma} \int_c^d {\tilde{\rho}_{(1)}^{{\rm MP}_\gamma}(x) \over \lambda/ n - x} \, dx
\end{equation}
where
$$
\tilde{\rho}_{(1)}^{{\rm MP}_\gamma}(x) := {\sqrt{(x-c)(d-x)} \over 2 \pi x}.
$$
Proceeding as in the analysis of the integral in (\ref{3.dyb1}), for $z := \lambda/n > d$ we write
\begin{equation}\label{s2.1}
 \int_c^d {\tilde{\rho}_{(1)}^{{\rm MP}_\gamma}(x) \over z - x} \, dx =
{1 \over z} \sum_{k=0}^\infty {1 \over z^k}
\int_c^d \tilde{\rho}_{(1)}^{{\rm MP}_\gamma}(x) x^k \, dx.
\end{equation}
Use can now be made of the known fact (see e.g.~\cite{Wa78})
\begin{equation}\label{s2.2}
\int_c^d \tilde{\rho}_{(1)}^{{\rm MP}_\gamma}(x) x^k \, dx =
\sum_{i=1}^k {1 \over k} \Big ( {k \atop i} \Big )
\Big ( {k \atop i-1} \Big ) \gamma^i,
\end{equation}
where the coefficient of $\gamma^i$ is the Narayana number $N(k,i-1)$.
Defining
$$
A_k(p,q) := \sum_{i=1}^k N(k,i-1) p^i q^{k+1-i}
$$
and introducing the generating function
$$
\psi(p,q,t) = \sum_{k=1}^\infty A_k(p,q) t^k
$$
it is  fundamental  to the theory of Narayana numbers \cite{St99} that
\begin{equation}\label{s2.3}
t \psi(p,q,t) = {1 \over 2} \Big ( 1 - u - v -
(1 - 2(u+v) + (u-v)^2)^{1/2} \Big )
\end{equation}
where $u:=pt$, $v:= qt$. Thus with (\ref{s2.2}) substituted in (\ref{s2.1}) the sum can be
expressed in the form (\ref{s2.3}) to give
\begin{equation}\label{s3.1}
 \int_c^d {\tilde{\rho}_{(1)}^{{\rm MP}_\gamma}(x) \over z - x} \, dx =
{1 \over 2} \Big ( 1 - {\gamma - 1 \over z} -
\Big ( 1 - {2(\gamma+1) \over z} + {(\gamma - 1)^2 \over z^2} \Big )^{1/2}
\Big ).
\end{equation}
Note that in the case $\gamma = 1$ we reclaim the evaluation of the integral implied by
(\ref{2.18}).

For $z := \lambda/n > d$ both terms in (\ref{s1.1}) are decreasing functions of $z$ and
so take on their maximum value when $z=d$. The square root in (\ref{s3.1}) then
vanishes, and we find that for (\ref{s1.1}) to have a solution we must have
$$
bn > \sqrt{\gamma} (1 +  \sqrt{\gamma} ).
$$
In this circumstance the value of $\lambda$ solving (\ref{s1.1}), and thus
the location of the separated eigenvalue can be made explicit.

\begin{prop}  \cite{BBP05} \label{pbib1}
Consider the rank 1 perturbation (\ref{fns}) and suppose $b \mapsto b/m$. In the limit $m \to
\infty$ with $n/m = \gamma \ge 1$ fixed, provided $b > 1 + 1/\sqrt{\gamma}$ a single
eigenvalue separates from the eigenvalue support in the case $b = 0$, and occurs at
$$
\lambda = b \Big ( 1 + {\gamma^{-1} \over b - 1} \Big ).
$$
\end{prop}

The reasoning that tells us Proposition \ref{pbib} holds in the real case implies too
that  Proposition \ref{pbib1} similarly holds in the real case.

It remains to study the eigenvalues of (\ref{3.1}) in the case that all entries of $X^{(0)}$
are equal to $\mu$. Thus
\begin{equation}\label{1.1a}
X^{(0)} = \mu \vec{1}_n \vec{1}_m^T
\end{equation}
where $\vec{1}_p$ is the $p \times 1$ vector with all entries equal to unity. In general
a matrix of the form (\ref{3.1}) has $n-m$ zero eigenvalues and $2m$ eigenvalues given
by $\pm$ the positive square roots of the eigenvalues of $ Y^\dagger Y$, 
$Y = X + X^{(0)}$. Correspondingly,
the matrix of eigenvectors has the structured form
\begin{equation}\label{UV}
\left [ \begin{array}{cc} U_{n \times n} & U_{n \times m} \\
V_{m \times n} & - V_{m \times m} \end{array} \right ]
\end{equation}
with the eigenvalues ordered from largest to smallest. Here $U_{n \times m}$
denotes the matrix $U_{n \times n}$ with the final $n - m$ columns deleted,
while $V_{m \times n}$ denotes $V_{m \times m}$ with $n-m$ columns of zeros added.
The matrix (\ref{UV}) is real orthogonal when (\ref{3.1}) has real entries, and is
unitary when the entries of (\ref{3.1}) are complex.

Substituting (\ref{1.1a}) for $X^{(0)}$ in (\ref{3.1}) and conjugating by (\ref{UV})
shows (\ref{3.1}) has the same eigenvalue distribution as
\begin{eqnarray}\label{duv}
&& {\rm diag} ( \lambda_1,\dots,\lambda_m, (0)^{n-m}, - \lambda, \dots, -
\lambda_m) \nonumber \\
&& \qquad + \mu
\left [ \begin{array}{cc} \vec{u}_n^* & \vec{v}_n^* \\
\vec{u}_m^* & -\vec{v}_m^* \end{array} \right ]
\left [ \begin{array}{cc} \vec{v}_n^T & - \vec{v}_m^T \\
\vec{u}_n^T & \vec{u}_m^T \end{array} \right ]
\end{eqnarray}
where here * denotes complex conjugation, and
\begin{eqnarray}\label{duvo}
\vec{u}_n^T := \vec{1}_n^T U_{n \times n}, \qquad
\vec{u}_m^T := \vec{1}_n^T U_{n \times m} \nonumber \\
\vec{v}_n^T := \vec{1}_m^T V_{m \times n}, \qquad
\vec{v}_m^T := \vec{1}_m^T V_{m \times m}.
\end{eqnarray}
With the diagonal matrix in (\ref{duv}) denoted $\Lambda$ the characteristic polynomial
for the matrix sum (\ref{duv}) can be written
\begin{equation}\label{3.1c}
\lambda^{n-m} \prod_{l=1}^m(\lambda^2 - \lambda_l^2)
\det \Big ( {\bf 1}_{n+m} + \mu
\left [ \begin{array}{cc} \vec{u}_n^* & \vec{v}_n^* \\
\vec{u}_m^* & -\vec{v}_m^* \end{array} \right ]
\left [ \begin{array}{cc} \vec{v}_n^T &  \vec{v}_m^Y \\
\vec{u}_n^T & -\vec{u}_m^T \end{array} \right ]
(\lambda  {\bf 1}_{n+m} - \Lambda)^{-1} \Big ).
\end{equation}
Use of the general formula
$$
\det ( {\bf 1}_p - A_{p \times q} B_{q \times p} ) =
\det ( {\bf 1}_q - B_{q \times p} A_{p \times q} ) 
$$
shows that the determinant in (\ref{3.1c}) is equal to the $2 \times 2$ determinant
\begin{equation}\label{3.1d}
\det \Big ( {\bf 1}_2 + \mu 
\left [ \begin{array}{cc} a_{11} & a_{12} \\
a_{21} & a_{22} \end{array} \right ] \Big )
\end{equation}
where, with $\vec{u}^T = (u_j)_{j=1,\dots,n}$ and $\vec{v}_n^T = ((v_j)_{j=1,\dots,m},(0)^{n-m})$,
\begin{eqnarray*}
&& a_{11} = \bar{a}_{22} = \sum_{j=1}^m {2 \lambda_j u_j \bar{v}_j \over \lambda^2 - \lambda_j^2} \\
&& a_{12} = \sum_{j=1}^m {2 \lambda |v_j|^2 \over \lambda^2 - \lambda_j^2 } + \sum_{j=m+1}^{n-m} {|v_j|^2 \over \lambda} \\ 
&& a_{21} = \sum_{j=1}^m {2 \lambda |u_j|^2 \over \lambda^2 - \lambda_j^2}
\end{eqnarray*} 

Substituting (\ref{3.1d}) in (\ref{3.1c}) shows (\ref{3.1}) again has $n-m$ zero
eigenvalues, with the remaining eigenvalues, which come in $\pm$ pairs, given by
the zeros of (\ref{3.1d}). From the definitions (\ref{duv}) and (\ref{duvo})
we deduce that
$$
\langle u_j \bar{v}_j \rangle = 0, \qquad 2 \langle |v_j|^2 \rangle =  2 \langle |u_j|^2 \rangle = 1 
$$
and so after averaging over the eigenvectors, and to leading order in $m$ with $n-m$ fixed,
the eigenvalue condition reduces to
\begin{equation}\label{3.1e}
1 = \mu \sum_{j=1}^m {\lambda \over \lambda^2 - \lambda_j^2}.
\end{equation}
The positive solutions of (\ref{3.1e}), $\{x_j\}_{j=1,\dots,m}$ say, exhibit the
interlacing
$$
0 < \lambda_1 < x_1 < \lambda_2 < x_2 < \cdots < \lambda_m < x_m
$$
and we know that the density of the positive eigenvalues is given by $2 \rho_{(1)}^{\rm W}(x)$
with 
\begin{equation}\label{2.6a}
N \mapsto m, \qquad J= 2 \sqrt{m}. 
\end{equation}
Hence, after averaging over $\{\lambda_j\}$ (\ref{3.1e}) becomes
the integral equation
\begin{equation}\label{3.1f}
1 = {\mu \lambda \over \pi} \int_0^1 {(1 - y^2)^{1/2} \over (\lambda^2/4m) - y^2 } \, dy.
\end{equation}
This integral is essentially the same as that appearing in (\ref{ie2}), so the evaluation
implied by (\ref{2.18}) allows the integral in (\ref{3.1f})
for $\lambda^2/4m > 1$ to be similarly
evaluated, giving
$$
1 = {\mu \lambda \over 2} \Big ( 1 - (1 - 4m/\lambda^2)^{1/2} \Big ).
$$
This is precisely the equation obtained by substituting (\ref{ie}) in
(\ref{3.dyb}), after the identifications (\ref{2.6a}). Hence we obtain
for the shifted mean chiral ensemble a result identical to that obtained
for the shifted mean Gaussian ensemble given in Proposition \ref{p.jk}.

\begin{prop}\label{psJ}
Consider shifted mean chiral ensemble matrices specified by (\ref{3.1}),
and set $\mu = c/J$, $J:= 2 \sqrt{m}$. For large $m$, with $n-m$ fixed
and $c>1$, a single eigenvalue separates from the support $(0,J)$ of the
positive eigenvalues, and its location is specified by (\ref{cc}).
\end{prop}

\section{Green function viewpoint}
The shifted mean Gaussian and shifted mean chiral random matrices 
can be used to formulate the Dyson Brownian motion of random matrices
(see e.g.~\cite{Fo02}). As we will see, this viewpoint can be used to exhibit 
the separation of a set of $r$ eigenvalues, in the setting that the matrices
$X^{(0)}$ in (\ref{1.1}) and (\ref{3.1}) have
a single non-zero eigenvalue and single non-zero singular value
$c$ say of degeneracy $r$, and that $c$ is large. Note that the
case $r=1$ corresponds to all entries being equal to $c/N$ and $c/m$ respectively.
Asymptotic formulas obtained in the course of so analyzing (\ref{1.1}) can then
be used to show an analogous effect for (\ref{1.2}).

Consider first (\ref{1.1}). We are interested in the p.d.f.~for the
eigenvalues $\{\lambda_j\}$ of $H$,
given the eigenvalues $\{\lambda_j^{(0)}\}$ of $H^{(0)}$. 
Let this be denoted $P^{\rm G}(\vec{\lambda}|\vec{\lambda}^{(0)})$. The Jacobian for the change
of variables from the entries of the matrix to the eigenvalues and eigenvectors is
proportional to
$$
\prod_{1 \le j < k \le N} |\lambda_k - \lambda_j|^\beta (U^\dagger dU)
$$
(see e.g.~\cite{Fo02}) where $(U^\dagger dU)$ is the normalized Haar measure for unitary matrices $(\beta = 2)$
and real orthogonal matrices $(\beta = 1)$. Hence the sought conditional eigenvalue p.d.f.~is
proportional to
\begin{equation}\label{tt.1}
\prod_{1 \le j < k \le N} |\lambda_k - \lambda_j|^\beta e^{-(\beta/2) \sum_{j=1}^N (\lambda_j^2 +
(\lambda_j^{(0)})^2)}
\int \exp \Big ( \beta {\rm Tr} (U \Lambda U^\dagger \Lambda^{(0)}) \Big )
(U^\dagger d U)
\end{equation}
where
$$
\Lambda = {\rm diag} (\lambda_1,\dots, \lambda_N), \qquad
\Lambda^{(0)}  = {\rm diag} (\lambda_1^{(0)},\dots, \lambda_N^{(0)}).
$$

Now, with $x:= \{ x_j \}_{j=1,\dots,n}$, $y := \{ y_j \}_{j=1,\dots,n}$ and
$C_\kappa^{(2/\beta)}(x)$ denoting the Schur ($\beta = 2$) or zonal ($\beta = 1$)
polynomial indexed by a partition $\kappa$ of $n$ parts, introduce the 
generalized hypergeometric function
\begin{equation}\label{tt.2.1}
{}_0^{} {\mathcal F}_0^{(\alpha)}(x;y) :=
\sum_{\kappa} {1 \over |\kappa|!}
{C_\kappa^{(\alpha)}(x) C_\kappa^{(\alpha)}(y) \over C_\kappa^{(\alpha)}((1)^n)}.
\end{equation}
It is a fundamental result in the theory of zonal polynomials that (\ref{tt.2.1}) relates
to (\ref{tt.1}) through the evaluation formula (see e.g.~\cite{Ma95})
\begin{equation}\label{tt.2.1a}
\int \exp \Big ( \beta {\rm Tr} (U \Lambda U^\dagger \Lambda^{(0)}) \Big )
(U^\dagger d U) = {}_0^{} {\mathcal F}_0^{(2/\beta)}(\beta \lambda^{(0)};\lambda)
\end{equation}
and thus the conditional eigenvalue p.d.f.~$P^{\rm G}(\vec{\lambda}|\vec{\lambda}^{(0)})$
is proportional to
\begin{equation}\label{tt.2.2}
\prod_{1 \le j < k \le N}|\lambda_k - \lambda_j|^\beta
e^{-(\beta/2) \sum_{j=1}^N (\lambda_j^2 + (\lambda_j^{(0)})^2)}
{}_0^{} {\mathcal F}_0^{(2/\beta)}(\beta \lambda^{(0)};\lambda).
\end{equation}

Generalizing (\ref{1.1}) so that it is now proportional to
\begin{equation}\label{ht}
\exp \Big ( - \beta {\rm Tr} ( H - e^{-\tau} H^{(0)})^2/2|1 - e^{-2\tau}| \Big )
\end{equation}
allows the corresponding eigenvalue probability density function for $H$, $p_\tau$ say
to be written as the solution of a Fokker-Planck equation. This is precisely the same
Fokker-Planck equation as appears in the Dyson Brownian motion model of the Gaussian
ensembles (see e.g.~\cite{Fo02}). It can equivalently be interpreted as the Brownian evolution
of a classical gas with potential energy
\begin{equation}\label{Wg}
W = - \sum_{1 \le j < k \le N} \log |\lambda_k - \lambda_j| + {1 \over 2} \sum_{j=1}^N \lambda_j^2
\end{equation}
is given explicitly by
\begin{equation}\label{fp}
{\partial p_\tau \over \partial \tau} = {\mathcal L} p_\tau \qquad {\rm where} \quad
{\mathcal L} = \sum_{j=1}^N {\partial \over \partial \lambda_j} \Big (
{\partial W \over \partial \lambda_j} + \beta^{-1} {\partial \over \partial \lambda_j} \Big ).
\end{equation}
The Green function solution $G_\tau^{\rm G}(\vec{\lambda}| \vec{\lambda}^{(0)})$ is that solution which satisfies the
initial condition
\begin{equation}\label{tt.pt}
p_\tau(\lambda) \Big |_{\tau = 0} = \prod_{l=1}^N \delta (\lambda_l - \lambda_l^{(0)})
\qquad (\lambda_N^{(0)} < \cdots < \lambda_1^{(0)}).
\end{equation}
It is given in terms of the generalized hypergeometric function (\ref{tt.2.1}) according to
\cite{BF97a}
\begin{eqnarray}\label{tt.3}
&& G_\tau^{\rm G}(\vec{\lambda}| \vec{\lambda}^{(0)}) = C e^{-\beta W(\lambda)} \\
\nonumber&& \qquad \times
\exp \Big ( - {\beta t^2 \over 2(1 - t^2)} \sum_{j=1}^N (\lambda_j^2 +
(\lambda_j^{(0)})^2 \Big )
{}_0^{} {\mathcal F}_0^{(2/\beta)} \Big ( {\beta \lambda t \over (1 - t^2)^{1/2}};
{ \lambda^{(0)} \over (1 - t^2)^{1/2} } \Big )
\end{eqnarray}
where $t := e^{-\tau}$ and here and below $C$ is {\it some} proportionality constant
independent of the primary quantities (here $\{\lambda_j\}$, $\{\lambda_j^{(0)}\}$) of the equation. 

Substituting (\ref{tt.3}) in (\ref{tt.2.1a}) allows the matrix integral to be
expressed in terms of the Green function. The significance of this for purposes of
studying the asymptotics of $P^{\rm G}(\vec{\lambda}|\vec{\lambda}^{(0)})$ in the
case that $\vec{\lambda}^{(0)}$ is given by
\begin{equation}\label{wr}
\vec{\lambda}^{(0)} = ((0)^{N-r}, (c)^r) 
\end{equation}
is that for $\tau \to 0$ the asymptotic form of $G_\tau^{\rm G}(\vec{\lambda}|\vec{\lambda}^{(0)})$
must factorize as
\begin{equation}\label{f1}
G_\tau^{\rm G}(\{\lambda_j\}_{j=r+1,\dots,N} | (0)^{N-r}) G_\tau^{\rm G}(\{\lambda_j\}_{j=1,\dots,r} | (c)^r )
\end{equation}
since the initial condition (\ref{tt.pt}) then separates the system into two. Furthermore, we know
from \cite{BF97a} that
\begin{equation}\label{f2}
G_\tau^{\rm G}(\{\lambda_j\}_{j=1,\dots,n} | (c)^n) \mathop{\sim}\limits_{\tau \to 0}
{\mathcal N}_{n,\beta,\tau} e^{- (\beta/4 \tau) \sum_{j=1}^n (\lambda_j - c)^2}
\prod_{1 \le j < k \le n} |\lambda_k - \lambda_j|^\beta.
\end{equation}
Making use of (\ref{f1}) and (\ref{f2}) in (\ref{tt.3}), and making use too of the scaling property 
\begin{equation}\label{tt.4}
{}_0^{} {\mathcal F}_0^{(2/\beta)} (x; ay) = {}_0^{} {\mathcal F}_0^{(2/\beta)} (ax; y),
\end{equation}
valid for $a$ scalar, allows the following asymptotic expansion to be deduced.

\begin{prop}\label{pF}
Let ${}_0 {\mathcal F}_0^{(2/\beta)}$ be specified by (\ref{tt.2.1}). 
In the limit $c \to \infty$
\begin{eqnarray}\label{bb.t}
&& \prod_{1 \le j < k \le N} |\lambda_k - \lambda_j |^\beta {}_0 {\mathcal F}_0^{(2/\beta)}(\beta \lambda;
((0)^{N-r},(c)^r)) \nonumber \\
&& \qquad \sim C e^{(\beta/2) \sum_{j=1}^r(\lambda_j^2 - (\lambda_j - c)^2 )}
\prod_{1 \le j < k \le r} |\lambda_k - \lambda_j|^\beta
\prod_{r+1 \le j < k \le N} |\lambda_k - \lambda_j|^\beta.
\end{eqnarray}
\end{prop}
For future reference we remark that the next order correction in (\ref{bb.t}) is to multiply the
right hand side by the factor (see e.g.~\cite{CP73})
\begin{equation}\label{bb.tp}
\prod_{j=1}^r \prod_{l=r+1}^N | \lambda_j - \lambda_l|^{\beta/2},
\end{equation}
which to leading order is a constant.

Substituting (\ref{bb.t}) in (\ref{tt.2.2}) exhibits the $c \to \infty$ form 
of $P^{\rm G}(\vec{\lambda}|\vec{\lambda}^{(0)})$.

\begin{cor}\label{pcF}
The conditional eigenvalue probability density $P^{\rm G}(\vec{\lambda}|\vec{\lambda}^{(0)})$, in the
case that $\vec{\lambda}^{(0)}$ is given by (\ref{wr}) with $c \to \infty$, factorizes to be proportional to
$$
\Big ( e^{- (\beta/2) \sum_{j=1}^r (\lambda_j - c)^2} \prod_{1 \le j < k \le r}
|\lambda_k - \lambda_j|^\beta \Big )
\Big ( e^{- (\beta/2) \sum_{j=r+1}^N \lambda_j^2} \prod_{r+1 \le j < k \le N}
|\lambda_k - \lambda_j|^\beta \Big ).
$$
We recognise the first term as the eigenvalue probability density function for the $r \times r$
Gaussian ensemble centred at $\lambda = c$, and the second term as the eigenvalue probability
density function for the $(N - r) \times (N - r)$ Gaussian ensemble centred at the origin.
\end{cor}

It is also possible to express the eigenvalue probability density function $P^{\rm L}$ for
the general variance Wishart matrices (\ref{1.2a}) in terms of the generalized hypergeometric
function ${}_0^{} {\mathcal F}^{(2/\beta)}_0$ \cite{Ja64}. We will revise this point, and
then proceed to make use of Proposition \ref{pF} to deduce a separation of eigenvalues in the
setting that the covariance matrix $\Sigma$ has $m-r$ eigenvalues equal to $1$ and $r$
eigenvalues equal to $b$, for $b \to \infty$. The case $r = 1$ corresponds to the setting
of Proposition \ref{pbib}.

To obtain the generalized hypergeometric function form of the eigenvalue probability density,
it is most convenient to consider as the input data not the eigenvalues of the covariance
matrix $\Sigma$, but rather its inverse $\Sigma^{-1}$. In particular, we are interested in the
case that
\begin{equation}\label{sS}
\Sigma^{-1} = {\rm diag} ( (\tilde{b})^r, (1)^{m-r}) := \Lambda^{(0)}.
\end{equation}
With the eigenvalues of (\ref{1.2a}) denoted $\{\lambda_j\}_{j=1,\dots,m}$ and
$\Lambda := {\rm diag} (\lambda_1,\dots,\lambda_m)$, by making use of relevant Jacobians
(see e.g.~\cite{Fo02}), $P^{\rm L}$ is seen to be proportional to
\begin{equation}\label{fn1.1}
\prod_{l=1}^m \lambda_l^{\alpha}  \prod_{1 \le j < k \le m} |\lambda_k - \lambda_j|^\beta
\int \exp \Big ( - {\beta \over 2} U \Lambda U^\dagger \Lambda^{(0)} \Big ) (U^\dagger dU), \quad
\alpha = (n-m+1/\beta)(\beta/2) - 1.
\end{equation}
The sought expression involving ${}_0^{} {\mathcal F}^{(2/\beta)}_0$ now follows by
substituting (\ref{tt.2.1a}), thus giving $P^{\rm L}$ as being proportional to
\begin{equation}\label{fn1.2}
\prod_{l=1}^m \lambda_l^{\alpha}  \prod_{1 \le j < k \le m} |\lambda_k - \lambda_j|^\beta
{}_0^{} {\mathcal F}^{(2/\beta)}_0(\lambda; - (\beta/2) \lambda^{(0)})
\end{equation}
where
\begin{equation}\label{fn1.3}
\lambda^{(0)} = ( (\tilde{b})^r,(1)^{m-r}).
\end{equation}

In (\ref{fn1.2}), the limit $\tilde{b} \to 0^+$ and thus $b \to \infty$ in the setting
of  Proposition \ref{pbib}, does not directly correspond to the setting of
Proposition \ref{pF}. However, if we first write $\lambda = u/\tilde{b}$, making use of (\ref{tt.4})
shows that we seek the asymptotics of
\begin{equation}\label{fn1.4}
{}_0^{} {\mathcal F}^{(2/\beta)}_0(u; - (\beta/2) ((1)^r, (1/\tilde{b})^{m-r})) =
e^{-(\beta/2) \sum_{j=1}^m u_j}
\, {}_0^{} {\mathcal F}^{(2/\beta)}_0(u; - (\beta/2) ((0)^r, (1/\tilde{b}-1)^{m-r}))
\end{equation}
where the equality follows from the matrix integral form of 
${}_0^{} {\mathcal F}^{(2/\beta)}_0$,
 (\ref{tt.2.1a}). We remark that writing $\lambda = u/\tilde{b}$ is well founded because for
 (\ref{1.2a}) all eigenvalues are positive, so we expect the variables $u$ to be $O(1)$
in the limit $\tilde{b} \to 0^+$.
Applying Proposition \ref{pbib}, modified to include the next order correction term
(\ref{bb.tp}), to (\ref{fn1.4}) and reverting back to the variables
$\lambda$ gives the asymptotic expansion 
\begin{eqnarray}\label{fn1.5}
&&\prod_{1 \le j < k \le m} |\lambda_k - \lambda_j|^\beta
{}_0^{} {\mathcal F}^{(2/\beta)}_0(\lambda; - (\beta/2) \lambda^{(0)}) 
\mathop{\sim}\limits_{\tilde{b} \to 0^+}
C e^{-(\beta/2) \tilde{b} \sum_{j=1}^r \lambda_j}
\nonumber \\
&& \qquad \times
e^{-(\beta/2) \sum_{j=r+1}^n \lambda_j}
\prod_{l=1}^r \lambda_l^{(\beta/2)(m-r)}
\prod_{1 \le j < k \le r} |\lambda_k - \lambda_j|^\beta
\prod_{r+1 \le j < k \le m} |\lambda_k - \lambda_j|^\beta.
\end{eqnarray}
Substituting this in (\ref{fn1.3}) we obtain the $\tilde{b} \to 0^+$ form of 
$P^{\rm L}$.

\begin{prop}\label{pLa}
The eigenvalue probability density $P^{\rm L}$ for general variance Wishart matrices
(\ref{1.2a}), in the case that $\Sigma^{-1}$ is given by (\ref{sS}), with
$\tilde{b} \to 0^+$,
factorizes to be proportional to
$$
\Big (\prod_{j=1}^r \lambda_j^{\alpha+\beta(m-r)/2} e^{-(\beta/2) \tilde{b} \lambda_j} \prod_{1 \le j < k \le r}
|\lambda_k - \lambda_j|^\beta \Big )
\Big (\prod_{j=r+1}^m \lambda^{\alpha}_j e^{-(\beta/2) \lambda_j}
 \prod_{r+1 \le j < k \le m}
|\lambda_k - \lambda_j|^\beta \Big ).
$$
We recognise the first term as the eigenvalue probability density function for the $r \times r$
Laguerre ensemble with $\lambda \mapsto \tilde{b} \lambda$, and $\alpha \mapsto \alpha + \beta (m-r)/2$,
and thus with support to leading
order at $O(1/\tilde{b})$, and the second term as the eigenvalue probability
density function for the $(m - r) \times (m - r)$ Laguerre ensemble.
\end{prop}

We turn our attention now to the shifted mean chiral matrices (\ref{3.1}), in the case that
$X^{(0)}$ has a single non-zero singular value $c$ of degeneracy $r$. As already noted, the
non-zero eigenvalues come in $\pm$ pairs, and there are $n-m$ zero eigenvalues. Furthermore,
the matrix of eigenvectors (\ref{UV}) has $U_{n \times n} =: U$ as the matrix of
eigenvectors of $Y Y^\dagger$, $Y := X + X^{(0)}$, and $V_{m \times m} := V$ the matrix
of eigenvectors of $Y^\dagger Y$. The Jacobian for the change of variables to the positive
eigenvalues and eigenvectors is proportional to
\begin{equation}\label{su.1}
\prod_{l=1}^m \lambda_l^{2 \alpha + 1}
\prod_{1 \le j < k \le m} |\lambda_k^2 - \lambda_j^2|^\beta
(U^\dagger dU) (V^\dagger dV)
\end{equation}
where $\alpha$ is as in (\ref{fn1.1}).

With $X$ distributed as a Gaussian matrix according to (\ref{1.2}) with $\Sigma = {\bf 1}_m$,
it follows from (\ref{su.1}) that the probability density function for the positive eigenvalues
of (\ref{3.1}) is proportional to
\begin{equation}\label{su.r21}
\prod_{l=1}^m \lambda_l^{2 \alpha + 1} e^{-(\beta/2) \lambda_l^2}
\prod_{1 \le j < k \le m} |\lambda_k^2 - \lambda_j^2|^\beta
\int (U^\dagger dU) \int (V^\dagger dV) \,
e^{\beta {\rm Re} \, {\rm Tr} (U \Lambda V \Lambda^{(0)\dagger})},
\end{equation}
where $\Lambda$ is the $n \times m$ matrix with all entries equal to 0 except those
on the diagonal which are equal to $\lambda_1,\dots\lambda_m$, and $\Lambda^{(0)\dagger}$
is the $m \times n$ matrix with all entries equal to zero except those in the first
$r$ positions of the diagonal, which are equal to $c$. We introduce now the further
generalized hypergeometric function
(\ref{tt.2.1})
\begin{equation}\label{su.2.1}
{}_0^{} {\mathcal F}_1^{(\alpha)}(a;x;y) :=
\sum_{\kappa} {[a]_\kappa^{(\alpha)} \over |\kappa|!}
{C_\kappa^{(\alpha)}(x) C_\kappa^{(\alpha)}(y) \over C_\kappa^{(\alpha)}((1)^n)},
\end{equation}
where $[a]_\kappa^{(\alpha)}$ is a certain generalized Pochammer symbol
(see e.g.~\cite{Fo02}). The matrix integral in (\ref{su.r21}) can be evaluated
in terms of ${}_0^{} {\mathcal F}_1^{(\alpha)}$ according to (see e.g.~\cite{Fo02})
\begin{equation}\label{su.2.2}
\int (U^\dagger dU) \int (V^\dagger dV) \,
e^{\beta {\rm Re} \, {\rm Tr} (U \Lambda V \Lambda^{(0)\dagger})}
= {}_0^{} {\mathcal F}_1^{(2/\beta)}(\beta n/2;\lambda^2;(\beta/2)^2(\lambda^{(0)})^2).
\end{equation}
Hence the  probability density function for the positive eigenvalues
of (\ref{3.1}) can be written in terms of ${}_0^{} {\mathcal F}_1^{(2/\beta)}$,
being proportional to
\begin{equation}\label{su.2.2a}
\prod_{l=1}^m \lambda_l^{2 \alpha + 1} e^{-(\beta/2) \lambda_l^2}
\prod_{1 \le j < k \le m} |\lambda_k^2 - \lambda_j^2|^\beta
 {}_0^{} {\mathcal F}_1^{(2/\beta)}(\beta n/2;\lambda^2;(\beta/2)^2(\lambda^{(0)})^2).
\end{equation}
 
Suppose now that the distribution of $Y := X + X^{(0)}$ is generalized to take on
a parameter dependent form proportional to
\begin{equation}\label{su.2.3}
\exp \Big ( - \beta {\rm Tr} ( Y - e^{-\tau} X^{(0)})^\dagger
( Y - e^{-\tau} X^{(0)})/2|1 - e^{-2\tau}| \Big )
\end{equation}
(cf.~(\ref{ht})). It is known (see e.g.~\cite{FNH99}) that the p.d.f.~of the positive
eigenvalues of $Y$, with those of $X^{(0)}$ regarded as given, satisfies the 
Fokker-Planck equation (\ref{fp}) with
\begin{equation}\label{10.28}
W = W^{\rm ch} = {1 \over 2} \sum_{j=1}^N x_j^2 - {\alpha' \over 2} \sum_{j=1}^N \log x_j^2 -
\sum_{1 \le j < k \le N} \log |x_k^2 - x_j^2|.
\end{equation}
Here $\alpha' = (2/\beta)(\alpha + 1/2)$ and $N=m$. This relates to the generalized
hypergeometric function (\ref{su.2.1}) through the fact that the Green function
solution of this Fokker-Planck equation, $G_\tau^{\rm ch}$ say, can be written \cite{BF97a}
\begin{eqnarray}\label{tt.3c}
&& G_\tau^{\rm ch}(\vec{\lambda}| \vec{\lambda}^{(0)}) = C e^{-\beta W^{\rm ch}(\lambda)} \\
\nonumber&& \qquad \times
\exp \Big ( - {\beta t \over 2(1 - t)} \sum_{j=1}^N (\lambda_j^2 +
(\lambda_j^{(0)})^2 \Big )
{}_0^{} {\mathcal F}_1^{(2/\beta)} \Big (\beta n/2;{\beta \lambda^2  \over 2 (1 - t)};
{\beta t (\lambda^{(0)})^2 \over 2 (1 - t)}  \Big )
\end{eqnarray}
where $t = e^{-2\tau}$ as in (\ref{tt.3}), and $\tilde{k}$ is independent of $\lambda, \lambda^{(0)}$.

We can view (\ref{tt.3c}) as allowing ${}_0^{} {\mathcal F}_1^{(2/\beta)}$ in (\ref{su.2.2a}) to be
rewritten in terms of the Green function. But as in (\ref{wr}), with 
\begin{equation}\label{wr1}
\vec{\lambda}^{(0)} = ((0)^{m-r}, (c)^r)
\end{equation}
the asymptotic form of $G_\tau^{\rm ch}(\vec{\lambda}|\vec{\lambda}^{(0)})$
must, for $\tau \to 0$, 
 factorize as
\begin{equation}\label{f1c}
G_\tau^{\rm ch}(\{\lambda_j\}_{j=r+1,\dots,m} | (0)^{m-r}) G_\tau^{\rm ch}(\{\lambda_j\}_{j=1,\dots,r} | (c)^r ).
\end{equation}
Furthermore, we know
from \cite{BF97a} that
\begin{equation}\label{f2c}
G_\tau^{\rm ch}(\{\lambda_j\}_{j=1,\dots,n} | (0)^n) \mathop{\sim}\limits_{\tau \to 0}
C \prod_{j=1}^n \lambda_j^{\beta \alpha'} e^{- (\beta/4 \tau) \sum_{j=1}^n \lambda_j^2}
\prod_{1 \le j < k \le n} |\lambda_k^2 - \lambda_j^2|^\beta,
\end{equation}
while for large $c$, the fact that with $x_j \mapsto \lambda_j + c$ (\ref{10.28})
reduces to (\ref{Wg}) tells us that 
\begin{equation}\label{f3}
G^{\rm ch}_\tau(\{\lambda\}_{j=1,\dots,r} |(c)^r) \: \sim \: 
G^{\rm G}_\tau(\{\lambda\}_{j=1,\dots,r} |(c)^r)
\end{equation}
and is thus given by (\ref{f2}). This enables us to deduce the
following asymptotic expansion, and then proceed to deduce the $c \to \infty$
form of $P^{\rm ch}(\vec{\lambda}|\vec{\lambda}^{(0)})$.

\begin{prop}\label{pFc}
Let ${}_0 {\mathcal F}_1^{(2/\beta)}$ be specified by (\ref{su.2.1}). 
In the limit $c \to \infty$
\begin{eqnarray}\label{bb.tc}
&& \prod_{l=1}^N \lambda_l^{\beta \alpha'}
\prod_{1 \le j < k \le N} |\lambda_k^2 - \lambda_j^2 |^\beta {}_0 {\mathcal F}_1^{(2/\beta)}(\beta n/2;
\lambda^2;
(\beta/2)^2 ((0)^{N-r},(c)^r)) \nonumber \\
&& \qquad \sim C e^{(\beta/2) \sum_{j=1}^r(\lambda_j^2 - (\lambda_j - c)^2 )}
\prod_{1 \le j < k \le r} |\lambda_k^2 - \lambda_j^2|^\beta
\prod_{l=r+1}^N \lambda_l^{\beta \alpha'}
\prod_{r+1 \le j < k \le N} |\lambda_k^2 - \lambda_j^2|^\beta
\end{eqnarray}
\end{prop}

\begin{cor}\label{pfr}
The conditional eigenvalue probability density $P^{\rm ch}(\vec{\lambda}|\vec{\lambda}^{(0)})$, in the
case that $\vec{\lambda}^{(0)}$ is given by (\ref{wr1}) with $c \to \infty$, factorizes to be proportional to
$$
\Big ( e^{- (\beta/2) \sum_{j=1}^r (\lambda_j - c)^2} \prod_{1 \le j < k \le r}
|\lambda_k - \lambda_j|^\beta \Big )
\Big ( \prod_{l=r+1}^m \lambda_l^{\beta \alpha'} e^{- (\beta/2) \lambda_l^2} \prod_{r+1 \le j < k \le m}
|\lambda_k^2 - \lambda_j^2 |^\beta \Big ).
$$
We recognise the first term as the eigenvalue probability density function for the $r \times r$
Gaussian ensemble centred at $\lambda = c$, and the second term as the eigenvalue probability
density function for the $(m - r) \times (m - r)$ chiral ensemble centred at the origin.
\end{cor}

\section{Explicit form of the correlations for $\beta = 2$}
According to (\ref{tt.1}) and (\ref{fn1.1}), the matrix integral in (\ref{tt.2.1a}) fully determines
the eigenvalue probability density function for the shifted mean Gaussian and general variance
Wishart ensembles. Furthermore, we know from (\ref{su.r21}) that the matrix integral in
(\ref{su.2.2}) fully determines the eigenvalue probability density function for the shifted
mean chiral ensemble. In the case $\beta = 2$ the matrix integrals in (\ref{fn1.1}) and
(\ref{su.2.2}) are over the Haar measure on the unitary group, and can be evaluated
in terms of determininants (see e.g.~\cite{Fo02}). In fact each of (\ref{tt.1}),
(\ref{fn1.1}) and  (\ref{su.r21}) can then be written as the product of two determinants.
From this functional form, using the general theory of biorthogonal ensembles
\cite{Bor99}, it is then possible to proceed to compute the general $n$-point
correlation functions $\rho_{(n)}$ as an $n \times n$ determinant,
\begin{equation}\label{rk}
\rho_{(n)}(\lambda_1,\dots,\lambda_n) = \det  [ K_N(\lambda_j,\lambda_k) ]_{j,k=1,\dots,n} 
\end{equation}
for a certain kernel function $K_N(x,y)$. In this section the eigenvalue separation
phenomenon for $\beta = 2$, exhibited at the level of the eigenvalue probability
density function in the previous section, will be analyzed both analytically and numerically
in terms of the correlation functions.

\begin{figure}
\begin{center}
\includegraphics{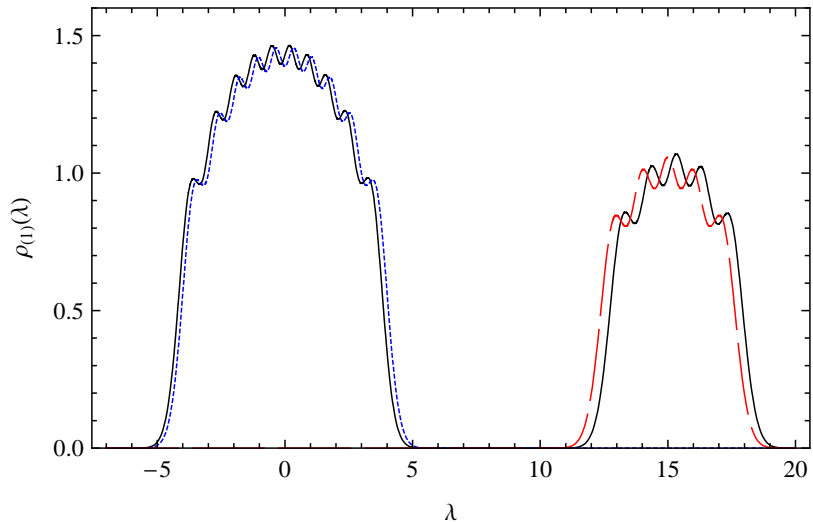}
\caption{(Color online) 
The eigenvalue probability density for the $N \times N$ GUE
for $N=15$ with $r=5$ shifted eigenvalues, 
$\lambda^{(0)}_i=c=15$, $i=1,\ldots,r$ (black solid line). 
The density
has two distinguishable parts, one is similar to the eigenvalue probability
density for the $(N-r) \times (N-r)$ GUE 
centered at the origin (blue dotted line) and the other is similar to the 
eigenvalue probability density for the 
$r \times r$ GUE
centered at $\lambda = c$ (red dashed line). These similarities become exact correspondences
in the limit $c \rightarrow \infty$.} 
\label{f.f1}
\end{center}
\end{figure}

Consider first the shifted mean Gaussian ensemble, and suppose that $X^{(0)}$ in  (\ref{tt.1})
has a single non-zero eigenvalue $c$ of degeneracy $r$, as assumed in Section 3.
Define
\begin{equation}\label{b.2.1}
K_n^{\rm GUE}(x,y) := {e^{-(x^2 + y^2)/2} \over \sqrt{\pi} }
\sum_{p=0}^{n-1} {H_p(x) H_p(y) \over 2^p p!},
\end{equation}
where $H_p(z)$ denotes the Hermite polynomial.
Then we know from \cite{DF06} that
\begin{equation}\label{b.2.2}
K_N(x,y) = K_{N-r}^{\rm GUE}(x,y) + \sum_{j=1}^r \tilde{\Gamma}^{(j)}(x) \Gamma^{(j)}(y),
\end{equation}
where, with ${\mathcal C}_{\{0,-2c\}}$ a simple closed contour encircling zero and $-2c$,
$\tilde{\Gamma}^{(j)}$ and $\Gamma^{(j)}$ are so called incomplete Hermite functions specified by
\begin{eqnarray}\label{b.2.3}
&& \tilde{\Gamma}^{(j)}(x) := \int_{{\mathcal C}_{\{0,-2c\}}} {e^{-xz - z^2/4} \over
z^{N-r} (z + 2c)^j } \, {dz \over 2 \pi i} \nonumber \\
&&  \Gamma^{(j)}(x) := \int_{-i \infty}^{i \infty} e^{xw + w^2/4} w^{N-r}
(w + 2c)^{j-1} \, {dw \over 2 \pi i}.
\end{eqnarray}

According to (\ref{rk}) setting $x=y$ in (\ref{b.2.2}) gives the eigenvalue density.
This being a function of one variable, it is well suited to a graphical representation.
Indeed the quantities (\ref{b.2.1}) and (\ref{b.2.3}) making up (\ref{b.2.2}) are all
readily computed numerically, so allowing for particular values of $N,r$ and $c$ the
density to be tabulated and then graphed. In regard to (\ref{b.2.3}), 
we first make use of the integral representations of the Hermite polynomial
\begin{eqnarray}\label{HnH}
H_n(x) & = & 2^n n! \int_{{\mathcal C}_{\{0\}}} {e^{x z - z^2/4} \over z^{n+1} } \, {dz \over 2 \pi i} \nonumber \\
& = & \sqrt{\pi} e^{x^2} \int_{-\infty}^\infty z^n e^{- xz + z^2/4} \, {dz \over 2 \pi i}
\end{eqnarray}
to evaluate the contour integrals. Consider for definiteness $\tilde{\Gamma}^{(1)}(x)$.
First computing the residue at $z = -2c$, then computing the contribution of the
singularity at the origin by writing
$$
{1 \over z + 2c} = {1 \over 2c} \sum_{p=0}^\infty (-1)^p {z^p \over (2c)^p },
$$
and using the first of the integral formulas in (\ref{HnH}) shows
\begin{equation}\label{Hnt2}
\tilde{\Gamma}^{(1)}(x) = (-1)^{N-1} \Big (
{e^{2cx - c^2} \over (2c)^{N-1} } - \sum_{p=0}^{N-2} {1 \over (2c)^{p+1} }
{H_{N-2-p}(x) \over 2^{N-2-p} (N-2-p)!} \Big ).
\end{equation}

\begin{figure}
\begin{center}
\includegraphics{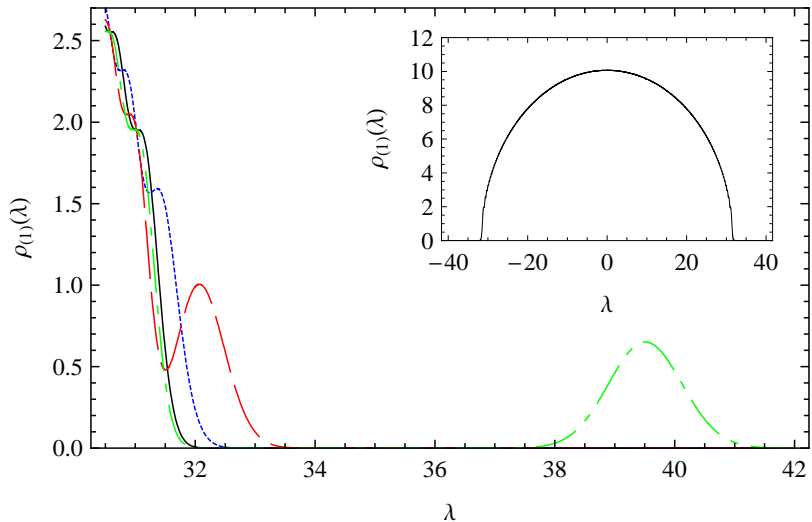}
\caption{(Color online) 
The onset of eigenvalue separation for
the $N \times N$ GUE
for $N=500$ with $r=1$ shifted eigenvalue, 
$\lambda^{(0)}_1=cJ/2$ and $J=\sqrt{2N}$.
Shown is the right edge of the eigenvalue probability density
for $c=0$ (black solid line), $c=1$ (blue dotted line),
$c=1.2$ (red dashed line), and $c=2$ (green dot-dashed line).
Inset shows the full eigenvalue probability density for the $N\times N$ 
unshifted mean GUE
with $N=500$.}  
\label{f.f1a}
\end{center}
\end{figure}

An example of a numerical calculation illustrating Corollary \ref{pcF}
is given in Figure \ref{f.f1}.
Another application is to use a numerical evaluation of (\ref{b.2.2}) to
illustrate the onset of the eigenvalue separation for large $N$ as
specified in Proposition \ref{p.jk}. This we do in Figure
\ref{f.f1a}. For these numerical calculations, we evaluated the integrals
in (\ref{b.2.3}) for arbitrary $j$ 
obtaining expressions involving sums of Hermite polynomials
as in (\ref{Hnt2}) and used them in (\ref{b.2.2}).

For general $x,y$ we can show analytically that the large $c$ form of $\rho_{(n)}$ is
consistent with Corollary \ref{pcF} by establishing the following asymptotic result.

\begin{prop}
For $c \to \infty$
\begin{equation}\label{b.3.1}
 \sum_{j=1}^r \tilde{\Gamma}^{(j)}(x) \Gamma^{(j)}(y) \: \sim \:
 e^{2c (x-y)} K_{r}^{\rm GUE}(x-c,y-c).
\end{equation}
\end{prop}

\noindent
{\it Proof}. \quad We see from the first formula in (\ref{b.2.3}) that
\begin{eqnarray*}
&&\tilde{\Gamma}^{(j)}(x)  \mathop{\sim}\limits_{c \to \infty}
\int_{{\mathcal C}_{\{-2c\}}} {e^{-xz - z^2/4} \over
z^{N-r} (z + 2c)^j } \, {dz \over 2 \pi i} = {e^{2cx - c^2} \over (-2c)^{N-r} }
\int_{{\mathcal C}_{\{-2c\}}} {e^{-(x-c)z - z^2/4} \over
z^j } \, {dz \over 2 \pi i} \nonumber \\
&& \qquad \qquad \qquad = {(-1)^{j-1} e^{2 c x - c^2} \over (-2c)^{N-r} }
{H_{j-1}(x-c) \over 2^{j-1} (j-1)!}
\end{eqnarray*}
where the second equality follows from the first integral formula in (\ref{HnH}). 
For the second integral formula in (\ref{b.2.3}), use of the second integral form
of the Hermite polynomials in (\ref{HnH}) shows
$$
\Gamma^{(j)}(y)  \mathop{\sim}\limits_{c \to \infty} (-1)^{j-1} (-2c)^{N-r} e^{-y^2}
{H_{j-1}(y-c) \over \sqrt{\pi} }.
$$
Substituting in the LHS of (\ref{b.3.1}) and comparing with (\ref{b.2.1}) gives
the RHS of (\ref{b.3.1}). \hfill $\square$

\medskip
According to (\ref{b.2.2}) and (\ref{b.3.1}), for $c \to \infty$ the correlation kernel $K_N$
is the sum of two terms, one with support in the neighbourhood of the origin, and the other
with support in the neighbourhood of $x,y = c$. Furthermore, these two terms are the correlation
kernel for the $(N -r) \times (N-r)$ GUE and the $r \times r$ GUE, the latter shifted
to be centred about $\lambda = c$ (note that the factor $e^{2c(x-y)}$ in (\ref{b.3.1}) does
note effect the determinant (\ref{rk})).

\begin{figure}
\begin{center}
\includegraphics{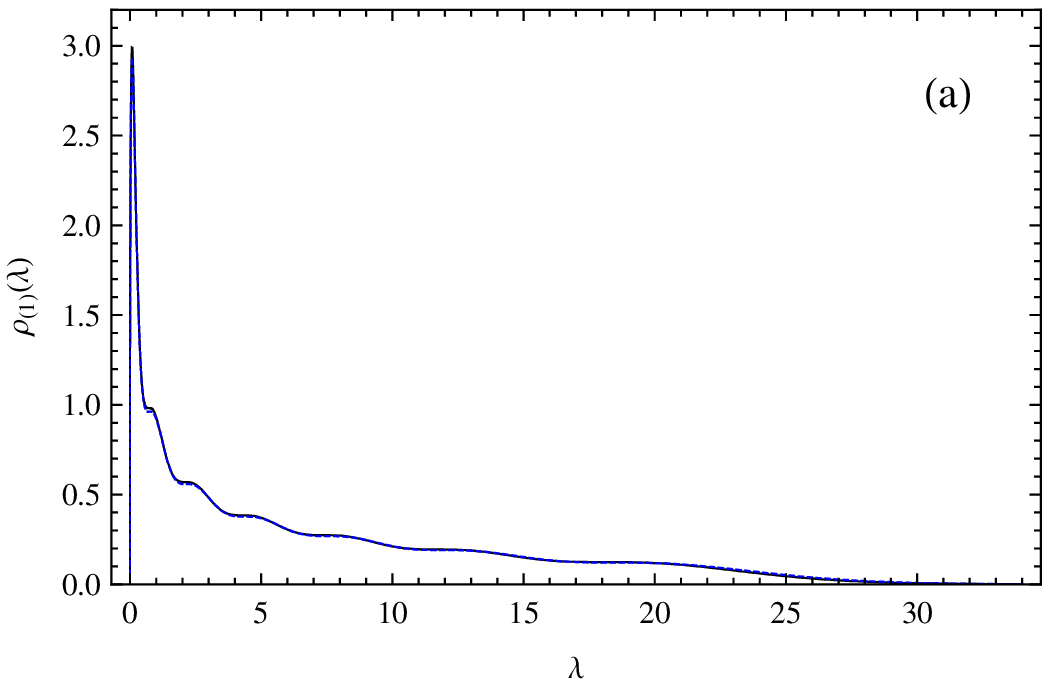}\\
\includegraphics{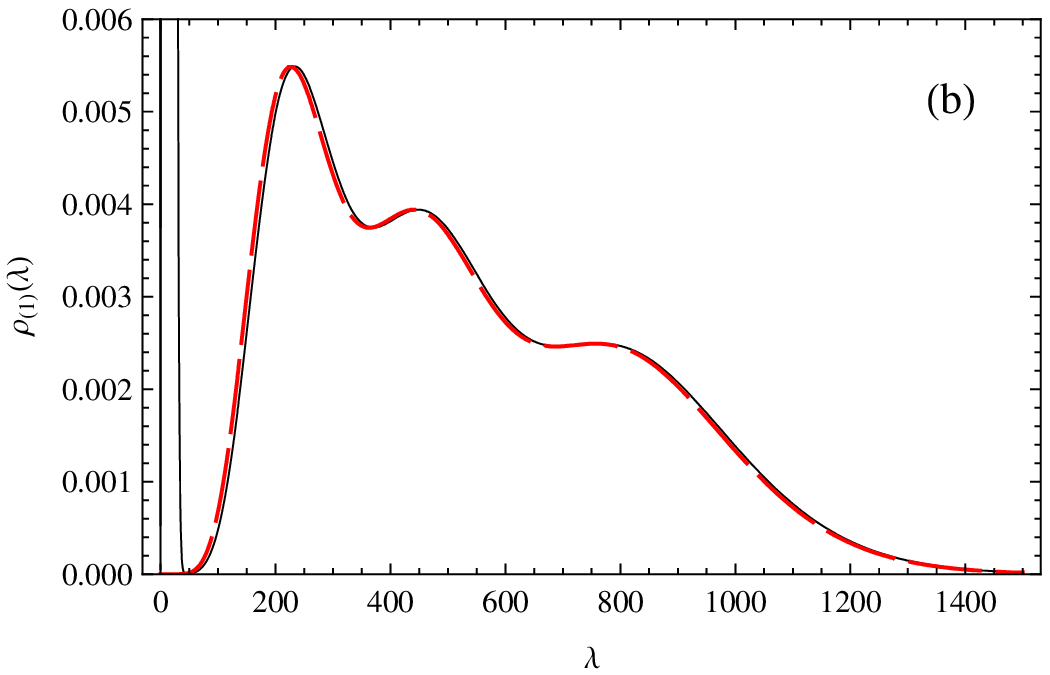}
\caption{(Color online) The eigenvalue probability density for the $m \times m$ 
variance LUE
or variance complex Wishart matrix ensemble
for $m=10$ and $\alpha = 1/2$ with $r=3$ shifted eigenvalues,
$\lambda^{(0)}_i=\tilde{b}=0.05$, $i=1,\ldots,r$ (black solid line). 
The density
has two distinguishable parts, one part, shown in $(a)$, is similar to the eigenvalue probability
density for the unshifted $(m-r) \times (m-r)$ variance LUE with unchanged $\alpha$
(blue dotted line) and the other part, shown in $(b)$, is similar to the 
eigenvalue probability density for the 
$r \times r$ variance LUE with $\lambda \mapsto \tilde{b} \lambda$ and
$\alpha \mapsto \alpha + (m-r)$ (thick red dashed line). These similarities become exact correspondences
in the limit $\tilde{b} \rightarrow 0^+$. Note that in $(a)$ the two lines (black solid and 
blue dotted) are indistinguishable and that the shifted part of the density is off-scale to
the right and not visible in the figure, while in $(b)$ a difference between the two 
lines (solid black and thick red dashed) is slightly apparent. The large
sharp peak at small values of
$\lambda$ in $(b)$ is the unshifted part the density shown in $(a)$.} 
\label{f.f2}
\end{center}
\end{figure}

In the case of  general variance complex Wishart matrices (\ref{1.2a}) with $\Sigma^{-1}$ specified
by (\ref{sS}), the correlations are given by (\ref{rk}) with $N=m$ and
\begin{equation}\label{dw.1}
K_m(x,y) = K_{m-r}^{\alpha + r}(x,y) + \sum_{i=1}^r \tilde{\Lambda}^{(i)}(x) \Lambda^{(i)}(y)
\end{equation}
(see \cite{DF06}). Here $\alpha = n - m$, and with $L_p^a(x)$ denoting the Laguerre polynomial
\begin{equation}\label{w.g}
K_n^a(x,y) = y^a e^{-y} \sum_{p=0}^{n-1} {(p+a)! \over p!} L_p^a(x) L_p^a(y)
\end{equation}
while $\tilde{\Lambda}^{(j)}$ and $\Lambda^{(j)}$ have been termed incomplete multiple
Laguerre functions and are specified by
\begin{eqnarray}\label{gg}
&& \tilde{\Lambda}^{(j)}(x) = \int_{{\mathcal C}_{\{0,(\tilde{b}-1)\}}}
{e^{-xz} (1 + z)^{m + \alpha} \over z^{m-r} (z - (\tilde{b}-1))^j } \, {dz \over 2 \pi i} \nonumber \\
&& {\Lambda}^{(j)}(x) = \int_{{\mathcal C}_{\{-1\}}}
{e^{xw} w^{m-r} (w -  (\tilde{b}-1))^{j-1} \over (1 + w)^{m+\alpha}  } \, {dw \over 2 \pi i}.
\end{eqnarray}

\begin{figure}
\begin{center}
\includegraphics{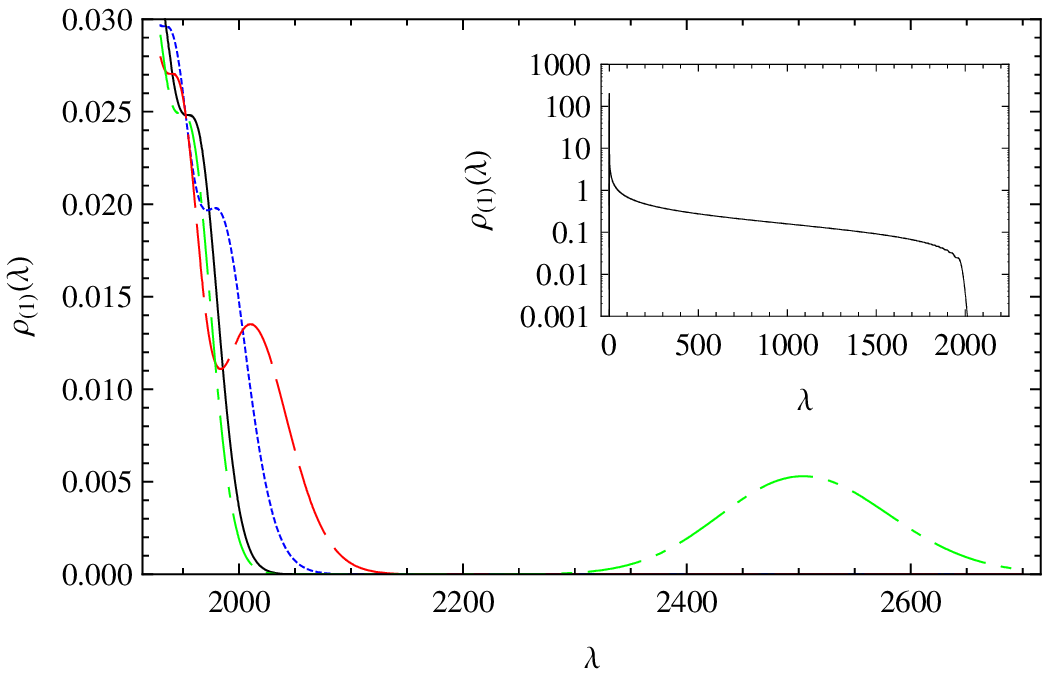}
\caption{(Color online) 
The onset of eigenvalue separation for
the $m \times m$ variance LUE 
or variance complex Wishart matrix ensemble
for $m=500$ and $\alpha=1/2$ with $r=1$ shifted eigenvalue, 
$\lambda^{(0)}_1=\tilde{b} J$ and $J=4m$.
Shown is the right edge of the eigenvalue probability density
for $\tilde{b}=0$ (black solid line), $\tilde{b}=0.5$ (blue dotted line),
$\tilde{b}=0.45$ (red dashed line), and $\tilde{b}=0.275$ (green dot-dashed line).
Inset shows the full eigenvalue probability density for the $m\times m$ unshifted mean
variance LUE
with $m=500$.
Note the logarithmic scale of the vertical axis of the inset graph.}  
\label{f.f2a}
\end{center}
\end{figure}

Making use of the integral representation for the Laguerre polynomials
\begin{equation}\label{L1}
L_n^a(x) = \int_{{\mathcal C}_{\{0\}}} {e^{-xw} \over w^{n+1} } (1 + w)^{n+a} \,
{dw \over 2 \pi i}
\end{equation}
allows $\tilde{\Lambda}^{(j)}(x)$ to be expressed in terms of these polynomials.
For example, by considering separately the neighbourhoods of zero and $\tilde{b}-1$ as
in the derivation of (\ref{Hnt2}), we can use (\ref{L1}) to show
$$
\tilde{\Lambda}^{(j)}(x) = {e^{-x(\tilde{b}-1)} \tilde{b}^{m+a} \over (\tilde{b} - 1)^{m-r} }
- \sum_{p=0}^{m-2} {L_{m-2-p}^{\alpha + p + 2}(x) \over (\tilde{b} - 1)^{p+1} }.
$$
To relate $\Lambda^{(j)}$ to Laguerre polynomials requires the identity
$$
n! (-x)^{-a} L_n^{-a}(x) = (n-a)! L_{n-a}^a(-x), \qquad a \in \mathbb Z,
$$
which when used in (\ref{L1}) implies the integral representation
$$
L_n^a(x) = {e^x x^{-a} (n+a)! \over n!}
\int_{{\mathcal C}_{\{-1\}}} {e^{xw} \over (w+1)^{n+a+1} } w^n \, {dw \over 2 \pi i}.
$$
From this we obtain, for example,
$$
\Lambda^{(1)}(x) = x^a e^{-x} L_{m-1}^a(x) {(m-1+a)! \over (m-1)!}.
$$

With the integrals (\ref{gg}) thus made explicit for arbitrary $j$, 
(\ref{dw.1}) in the case $x=y$, and
for particular $m,\alpha$ and $r$ can readily be tabulated and graphed. An example
illustrating Proposition \ref{pLa} is given in Figure \ref{f.f2}. 
However, unlike the situation with (\ref{b.2.2}), the structure (\ref{dw.1}) is not well
suited to exhibit the eigenvalue separation effect of Proposition \ref{pLa} in
a single plot. This is due to
the dependence on $\alpha+r$ rather than $\alpha$ in the first term on the RHS.
We can also make use of a numerical evaluation of (\ref{dw.1}) in
the case
$x=y$ to illustrate the onset of the eigenvalue separation for
large $m$
and $\tilde{b}$ near 1/2, as predicted by Proposition \ref{pbib}.
This is done in Figure \ref{f.f2a}.

\begin{figure}
\begin{center}
\includegraphics{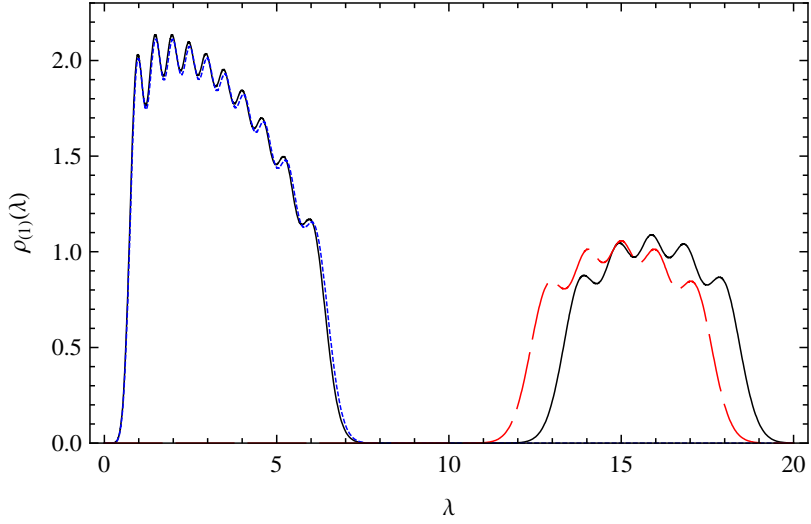}
\caption{(Color online) 
The eigenvalue probability density for the $m \times m$ 
chiral matrix ensemble
for $m=15$ and $\alpha = 4$ with $r=5$ shifted eigenvalues,
$\lambda^{(0)}_i=c=15$, $i=1,\ldots,r$ (black solid line). 
The density
has two distinguishable parts, one part is similar to the eigenvalue probability
density for the $(m-r) \times (m-r)$ unshifted mean chiral matrix ensemble
with the same $\alpha$
(blue dotted line) and the other part is similar to the 
eigenvalue probability density for the 
$r \times r$ GUE
centered at $\lambda = c$ (red dashed line). These similarities become exact correspondences
in the limit $c \rightarrow \infty$. 
} 
\label{f.f3}
\end{center}
\end{figure}

It remains to consider the shifted mean complex chiral matrices (\ref{3.1}) in the case that
$X^{(0)}$ has a single non-zero singular value $c$ of degeneracy $r$ as in Section 3.
According to \cite{DF07} the correlations are specified by
\begin{equation}\label{w.4.0}
\rho_{(l)}(x_1,\dots,x_l) = 2^l \prod_{\nu=1}^n x_\nu
\det [ K_m(x_j^2, x_k^2) ]_{j,k=1,\dots,l}
\end{equation}
where
\begin{equation}\label{w.4.1}
K_m(x,y) = K_{m-r}^\alpha(x,y) + \sum_{i=1}^r p_i(x) q_i(y).
\end{equation}
In (\ref{w.4.1}) $K_{m-r}^\alpha$ is specified by (\ref{w.g}), while
\begin{eqnarray}\label{w.4.2}
&& p_k(x) = {e^x \over \Gamma(\alpha + 1) } \int_0^\infty u^{m+\alpha - r}
(u + c^2)^{k-1} e^{-u} \,{}_0F_1(\alpha+1;-xu) \, du\\
&&q_k(x) = {x^\alpha e^{-x} \over \Gamma(\alpha + 1) }
\int_{{\mathcal C}_{\{-1,-c^2\}}}
{e^{v} {}_0F_1(\alpha+1;-xv) \over v^{m - r} (v + c^2)^{k}  } \, {dv \over 2 \pi i}.
\end{eqnarray}

The functions $p_k(x)$, $q_k(x)$ can be expressed in terms of Laguerre polynomials.
To see this, for $p_k(x)$ we require the integral formula 
\begin{equation}\label{w.4.3}
L_n^\alpha(x) = {e^x \over n! \Gamma(\alpha + 1) }
\int_0^\infty u^{\alpha + n} e^{-u} \, {}_0 F_1(\alpha + 1; - xu) \, du,
\end{equation}
while for $q_k(x)$ we require
\begin{equation}\label{w.5.1}
{L_n^k(x) \over \Gamma(n+k+1) } = {1 \over \Gamma(k+1) }
\int_{{\mathcal C}_{\{0\}}}
{e^w \, {}_0F_1(k+1;-xw) \over w^{n + 1} } \, {dw \over 2 \pi i}.
\end{equation}

\begin{figure}
\begin{center}
\includegraphics{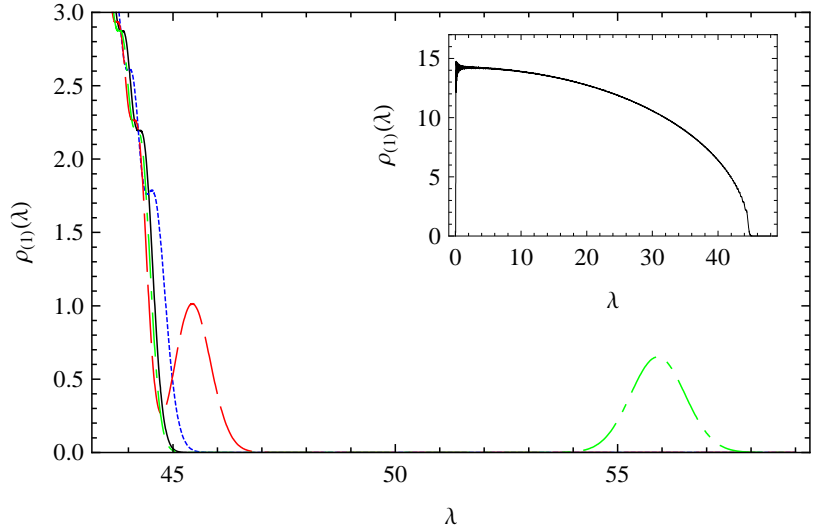}
\caption{(Color online) 
The onset of eigenvalue separation for
the $m \times m$ 
chiral matrix ensemble
for $m=500$ and $\alpha=2$ with $r=1$ shifted eigenvalue, 
$\lambda^{(0)}_1=cJ/2$ and $J=2\sqrt{m}$.
Shown is the right edge of the eigenvalue probability density
for $c=0$ (black solid line), $c=1$ (blue dotted line),
$c=1.2$ (red dashed line), and $c=2$ (green dot-dashed line).
Inset shows the full eigenvalue probability density for the $m\times m$ unshifted
mean chiral matrix ensemble
with $m=500$ and $\alpha=2$.}  
\label{f.f3a}
\end{center}
\end{figure}

Once $p_k(x), q_k(x)$ have been expressed in terms of Laguerre polynomials,
the kernel (\ref{w.4.1}) is for $x=y$ and particular $m, \alpha, r$ readily
computed numerically. A plot illustrating the eigenvalue separation property
Corollary \ref{pfr} is given in Figure \ref{f.f3}. 
A further plot illustrating the onset of the large $m$ separation
of a single eigenvalue, for appropriate $c$ as specified in
Proposition \ref{psJ}, is given in Figure \ref{f.f3a}.

The correlation (\ref{w.4.0})
for general $l$ can be shown analytically to be consistent with Corollary \ref{pfr}.
For this the $c \to \infty$ behaviour of the summation (\ref{w.4.1}) is required.

\begin{prop}
For large $c$
\begin{equation}\label{qk}
q_k(x) \sim {(-1)^{N-r+k-1} \over c^{2(N-r) + k} } 
{e^{-(\sqrt{x} - c)^2} \over \sqrt{\pi} 2^k (k-1)! } H_{k-1}(\sqrt{x} - c)
\end{equation}
while for large $c$ with $\sqrt{x} - c$ fixed 
\begin{equation}\label{pk}
p_k(x) \sim (-1)^{N-r+k-1} c^{2(N-r) + k-1}  H_{k-1}(\sqrt{x} - c).
\end{equation}
Hence for large $c$
\begin{equation}\label{pqk}
 \sum_{i=1}^r p_i(x) q_i(y) \sim {e^{-(\sqrt{y}-c)^2/2 + (\sqrt{x}-c)^2/2} \over 2 c} K_r^{\rm GUE}(\sqrt{x}-c,\sqrt{y}-c).
\end{equation}
\end{prop}

\noindent
{\it Proof.} \quad We require the formula
\begin{equation}\label{If1}
I_\alpha(2 \sqrt{z}) = {z^{\alpha/2} \over \Gamma(\alpha + 1) } \, {}_0 F_1(\alpha+1;z)
\end{equation}
and the asymptotic expansion
\begin{equation}\label{If2}
I_\alpha(x) \mathop{\sim}\limits_{x \to 0} {e^x \over (2 \pi x)^{1/2} }.
\end{equation}
Consider first $q_k(x)$. For $c$ large we see that the main contribution comes from the
singularity at $v = - c^2$. Writing $v \mapsto - c^2 + v$, making use of (\ref{If1}) and
(\ref{If2}), then expanding the exponent to second order in $v$ while keeping only the 
leading term in the rest of the integrand shows
$$
q_k(x) \sim {c^{-\alpha} x^{\alpha/2} e^{-(\sqrt{x} - c)^2} \over
2 \sqrt{\pi} (\sqrt{x c^2})^{1/2} (-c^2)^{N-r} } {1 \over c^{k-1} }
\int_{{\mathcal C}_{\{0\}}} {e^{- (\sqrt{x} - c) v - v^2/4} \over v^k } \,
{dv \over 2 \pi i}.
$$
The contour integral can be evaluated in terms of a Hermite polynomial according to
the first formula in (\ref{HnH}), and furthermore $x$ can be replaced by
$c^2$ in the prefactors not involving
the difference $\sqrt{x} - c$. This gives (\ref{qk}). 

Consider now $p_k(x)$. Making use of (\ref{If1}) and (\ref{If2}), then
expanding the exponent to second order about the stationary point at $u=-x$
and deforming the contour into the direction of steepest descents shows
$$
p_k(x) \sim (-1)^{N-r} {c^{2(N-r) + k - 1} \over 2 \sqrt{\pi} }
\int_{-\infty}^\infty (iu - 2(\sqrt{x} - c) )^{k-1} e^{ - u^2/4} \, du.
$$
Recalling now the second integral formula for the Hermite polynomials in
(\ref{HnH}) gives (\ref{pk}). Substituting (\ref{qk}) and (\ref{pk})
in the LHS of (\ref{pqk}), and recalling (\ref{b.2.1}) gives the RHS of
(\ref{pqk}). \hfill $\square$

\medskip
Substituting (\ref{pqk}) in (\ref{w.4.0}) we see that for $c \to \infty$ the
correlation kernel separates into two parts, and that the resulting correlations
are those corresponding to the two ensembles identified in Corollary \ref{pfr}.

\begin{figure}
\begin{center}
\includegraphics{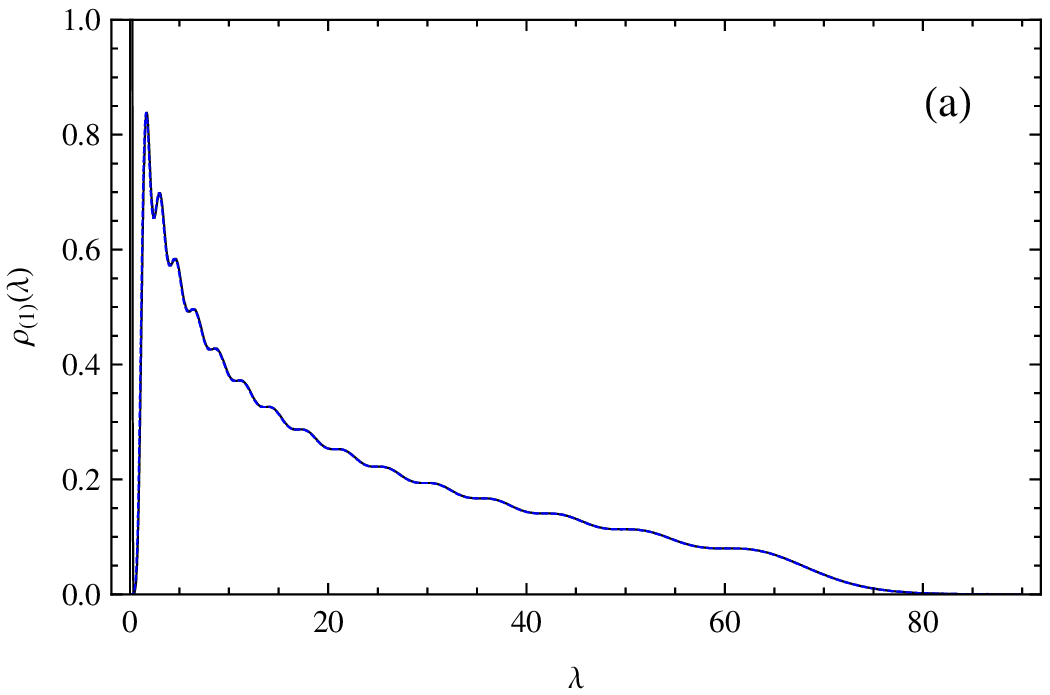}\\
\includegraphics{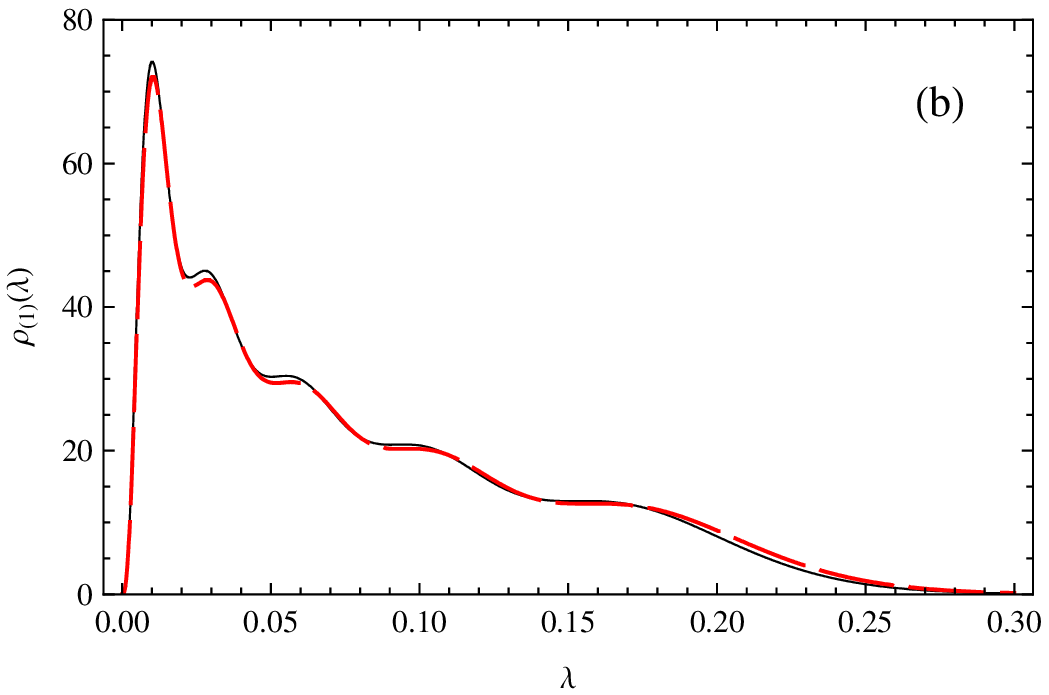}
\caption{(Color online) The eigenvalue probability density for the $m \times m$ 
variance LUE
or variance complex Wishart matrix ensemble
for $m=20$ and $\alpha = 3$ with $r=5$ shifted eigenvalues,
$\lambda^{(0)}_i=\tilde{b}=100$, $i=1,\ldots,r$ (black solid line). 
The density
has two distinguishable parts, one part, shown in $(a)$, is similar to the eigenvalue probability
density for the unshifted $(m-r) \times (m-r)$ variance LUE with $\alpha \mapsto \alpha + r$
(blue dotted line) and the other part, shown in $(b)$, is similar to the 
eigenvalue probability density for the 
$r \times r$ variance LUE with $\lambda \mapsto \tilde{b} \lambda$ and
unchanged $\alpha$ (thick red dashed line). These similarities become exact correspondences
in the limit $\tilde{b} \rightarrow \infty$. Note that in $(a)$ the two lines (black solid and 
blue dotted) are almost indistinguishable and that 
the large
sharp peak at small values of
$\lambda$, which does not correspond to the blue dotted line, 
is the shifted part the density shown in $(b)$.
In $(b)$
the unshifted part of the density is off-scale to
the right and not visible in the figure.
}
\label{f.f4}
\end{center}
\end{figure}

\section{A further asymptotic limit for the general variance Laguerre ensemble}
We see from (\ref{fn1.2}) that the general variance Wishart matrices (\ref{1.2a}),
in the case that $\Sigma^{-1}$ is specified by (\ref{sS}), is well suited to asymptotic
analysis of the $\tilde{b} \to \infty$ limit. In the setting of Proposition \ref{pbib}
this corresponds to $b \to 0$, so the analysis leading to the eigenvalue separation
therein has no bearing to this limit. Starting with (\ref{dw.1}), and with $m \to \infty$,
the scaled correlation functions corresponding to this limit were analyzed in
\cite{DF06}. The scaling was specified by setting $x_i \mapsto x_i/m$ and
$\tilde{b} = m c$, $c$ fixed. Here we will consider the limit $\tilde{b} \to
\infty$ with $m$ fixed.

\begin{prop}
The eigenvalue probability density $P^{\rm L}$ for general variance Wishart
matrices in the case that $\Sigma^{-1}$ is given by (\ref{sS}), with
$\tilde{b} \to \infty$,
factorizes to be proportional to
$$
\Big (\prod_{j=1}^r \lambda_j^{\alpha} e^{-(\beta/2) \tilde{b} \lambda_j} \prod_{1 \le j < k \le r}
|\lambda_k - \lambda_j|^\beta \Big )
\Big (\prod_{j=r+1}^m \lambda^{\alpha+r}_j e^{-(\beta/2) \lambda_j}
 \prod_{r+1 \le j < k \le m}
|\lambda_k - \lambda_j|^\beta \Big ).
$$
We recognise the first term as the eigenvalue probability density function for the $r \times r$
Laguerre ensemble with $\lambda \mapsto \tilde{b} \lambda$
and thus with support to leading
order at $O(1/\tilde{b})$, and the second term as the eigenvalue probability
density function for the $(m - r) \times (m - r)$ Laguerre ensemble with
$\alpha \mapsto \alpha + r$.
\end{prop}

\noindent
{\it Proof.} \quad This is an immediate consequence of the asymptotic formula
(\ref{fn1.4}), modified to include the correction term (\ref{bb.tp}).
\hfill $\square$

\medskip
In the case $\beta = 2$ the eigenvalue separation property can be exhibited 
numerically using (\ref{dw.1}). For this we proceed as in the lead up to
Figure \ref{f.f2}, making use of evaluations in terms of Laguerre polynomials
of the quantities (\ref{gg}). An example is given in Figure \ref{f.f4}.

\section{Concluding remarks}
The prime motivation for our work was to advance the understanding
of the precise effect shifting the mean in the Gaussian probability
distribution for the elements (\ref{1.1}) has on the eigenvalue 
distribution.
We sought to compare this to analogous
effects for general variance Wishart matrices, and shifted
mean chiral ensembles. We also had in mind the application of our work
in the fields of biological webs, food chains, plant and animal
ecology, and networks, neural and otherwise.

During the period of the 1950s through the 1970s there was an intense
activity by ecologists in studying the role played by biodiversity on
ecological systems. In particular, how the number of species of a
community affected its stability. There was quite conflicting evidence
as to whether complexity increased or decreased or had no effect on
the stability of an ecosystem.

The {\it coup de grace} came with the classic work of May
\cite{i,ii,iii} from purely a theoretical approach. This work has inspired
literally a vast amount of literature on the complexity-stability
issue in these and other arenas. May used the notions and technical
details from the mathematics of random matrix theory in formulating his
simple as he called them mathematical models with their predication
that increasing complexity decreases the stability of the system.

This pioneering work has motivated in the past 35 years and will
continue to do so into the future the debate on the complexity-
stability issue. In particular, it has been evidenced that under
certain specific conditions in the system that complexity can either
increase or decrease the system's stability. In an interesting toy
model \cite{iv} the authors have studied an ecological system where they use
a small matrix where the elements are taken from a random distribution
of interaction coefficients. They found that the non zero mean
of the distribution can lead to either an increase or decrease in
the stability of the system. (See also some very early work of 1971
\cite{v} discussed by May \cite{ii,iii}.)

Although May's work was strongly influenced by random matrix theory,
there has been little application of the mathematics
of random matrix theory to these vitally interesting areas and issues.
Specifically, it is the distribution of the largest eigenvalue(s) of a
random matrix that determines its stability. As we have shown in this
paper, a non zero mean in the probability distribution for the
elements of a random matrix severely affects its eigenvalue distribution
and hence its stability.

Very recent works \cite{vi,vii,viii,BS08} studying the eigenvalues
of the real Ginibre ensemble \cite{ix} have strong potential bearing on
the further application to biological webs and neural networks
\cite{x,xi,xii}. The generalization to the case of
complex eigenvalues is very important
and we plan to study this in our future work.

\section*{Acknowledgements}
KEB was supported by the NSF through grant \#DMR-0427538 and by the
Texas Advanced Research Program through grant \#95921. NEF appreciates
the support received during a visit early this year from this
NSF grant, and from the M-OSRP Center of Professor Arthur B.\ Weglein,
who he gratefully thanks for everything. PJF was supported by the
Australian Research Council.


\begin{thebibliography}{10}

\bibitem{ADV07}
M.~Adler, J.~Del\'epine, and P.~van Moerbeke, \emph{Dyson's non-intersecting
  {B}rownian motions with a few outliers}, arXiv:0707.0442, 2007.

\bibitem{viii}
G. Akemann and E. Kanzieper, \emph{Integrable structure of Ginibre's
ensemble of real random matrices and Pfaffian integration
theorem}, J.Stat.Phys. \textbf{129} (2007), 1159-1281.

\bibitem{An96}
J.~Anderson, \emph{A secular equation for the eigenvalues of a diagonal matrix
  perturbation}, Linear Alg. Applications \textbf{246} (1996), 49--70.

\bibitem{BY08}
Z.~Bai and J.-F. Yao, \emph{Central limit theorems for eigenvalues in a spiked
  population model}, Ann. l'Institut H. Poicar\'e \textbf{44} (2008), 447--474.

\bibitem{BBP05}
J.~Baik, G.~Ben Arous, and S.~P\'ech\'e, \emph{Phase transition of the largest
  eigenvalue for nonnull complex sample covariance matrices}, Annals of Prob.
  \textbf{33} (2005), 1643--1697.

\bibitem{BS06}
J.~Baik and J.W. Silverstein, \emph{Eigenvalues of large sample covariance
  matrices of spiked population models}, J. Mult. Anal. \textbf{97} (2006),
  1382--1408.

\bibitem{BF97a}
T.H. Baker and P.J. Forrester, \emph{The {Calogero-Sutherland} model and
  generalized classical polynomials}, Commun. Math. Phys. \textbf{188} (1997),
  175--216.

\bibitem{BL08}
M.~Bertola and S.Y. Lee, \emph{First colonization of a hard edge in random
  matrix theory}, arXiv:0804.1111, 2008.

\bibitem{BK04a}
P.M. Bleher and A.~Kuijlaars, \emph{Large $n$ limit of {G}aussian random
  matrices with external sources {I}}, Comm. Math. Phys. \textbf{252} (2004),
  43--76.

\bibitem{Bor99}
A.~Borodin, \emph{Biorthogonal ensembles}, Nucl. Phys. B \textbf{536} (1998),
  704--732, 2008.

\bibitem{BS08}
A.~Borodin and C.D. Sinclair, \emph{The Ginibre ensemble of real random matrices and
its scaling limits}, arXiv:0805.2986v1

\bibitem{CP73}
A.K. Chattopadhyay and K.C.S. Pillai, \emph{Asymptotic expansions for the
  distributions of characteristic roots when the parameter matrix has several
  multiple roots}, Multivariate Analysis III (P.R. Krishnaiah, ed.), Academic
  Press, New York, 1973, pp.~117--127.

\bibitem{DF06}
P.~Desrosiers and P.J. Forrester, \emph{Hermite and {L}aguerre
  $\beta$-ensembles: asymptotic corrections to the eigenvalue density}, Nucl.
  Phys. B \textbf{743} (2006), 307--332.

\bibitem{DF07}
\bysame, \emph{A note on biorthogonal ensembles}, J. Approx. Th. \textbf{152}
  (2007), 167--187.

\bibitem{Dy62b}
F.J. Dyson, \emph{A {B}rownian motion model for the eigenvalues of a random
  matrix}, J. Math. Phys. \textbf{3} (1962), 1191--1198.

\bibitem{xi}
J.Feng, V.K. Jirsa, and M.Ding, \emph{Synchronization in networks
with random interactions: Theory and applications}, Chaos \textbf{16}
(2006), 015019.

\bibitem{Fo02}
P.J. Forrester, \emph{Log-gases and {Random} {Matrices}},
  www.ms.unimelb.edu.au/\~{}matpjf/matpjf.html.

\bibitem{vi}
P.J. Forrester and T. Nagao, \emph{Eigenvalue statistics of the
real Ginibre Ensemble}, Phys. Rev. Lett. \textbf{99}, (2007) 050603.

\bibitem{FNH99}
P.J. Forrester, T.~Nagao, and G.~Honner, \emph{Correlations for the
  orthogonal-unitary and symplectic-unitary transitions at the hard and soft
  edges}, Nucl. Phys. B \textbf{553} (1999), 601--643.

\bibitem{FK81}
Z.~Furedi and J.~Komlos, \emph{The eigenvalues of random symmetric  
matrices}, Combinatorica \textbf{1} (1981),
  233--241.

\bibitem{ix}
J. Ginibre, \emph{Statistical ensembles of complex, quaternion,
and real matrices}, J. Math. Phys. \textbf{6}, (1965) 440-449.

\bibitem{IS05}
T.~Imamura and T.~Sasamoto, \emph{Polynuclear growth model, GOE${}^2$ and
  random matrix with deterministic source}, Phys. Rev. E \textbf{71} (2005),
  041606.

\bibitem{Ja64}
A.T. James, \emph{Distributions of matrix variates and latent roots derived from
  normal samples}, Ann. Math. Statist. \textbf{35} (1964), 475--501.

\bibitem{iv} 
V.A.A. Jansen and G.D. Kokoris, \emph{Complexity and
stability revisited}, Ecology Letters \textbf{6}, (2003),  498--502.

\bibitem{JKT78}
R.C. Jones, J.M. Kosterlitz, and D.J. Thouless, \emph{The eigenvalue spectrum
  of a large symmetric random matrix with a finite mean}, J. Phys. A
  \textbf{11} (1978), L45--L48.

\bibitem{KTJ78}
J.M. Kosterlitz, D.J. Thouless, and R.C. Jones, \emph{Spherical model of a spin
  glass}, Phys. Rev. Lett. \textbf{36} (1976), 1217--1220.

\bibitem{La64}
D.W. Lang, \emph{Isolated eigenvalue of a random matrix}, Phys. Rev.
  \textbf{135} (1964), B1082--B1084.

\bibitem{Ma95}
I.G. Macdonald, \emph{Hall polynomials and symmetric functions}, 2nd ed.,
  Oxford University Press, Oxford, 1995.

\bibitem{Ma07}
M.~Maida, \emph{Large deviations for the largest eigenvalue of rank one
  deformations of Gaussian ensembles}, Elec. J. Prob. \textbf{12} (2007),
  1131--1150.

\bibitem{i}
R.M. May, \emph{Will a large complex system be stable?}, 
 Nature \textbf{238} (1972), 413--414.

\bibitem{ii}
\bysame, \emph{Stability and complexity in model ecosystems},
Princeton University Press, Princeton, 1973.

\bibitem{iii}
\bysame, \emph{Some mathematical questions in biology},
edited
by J.D. Cowan, American Mathematical Society, Washington, DC,
1974.

\bibitem{v}
S. Makridakis and R. Weintraub, \emph{On the synthesis of general
systems}, General System \textbf{16}, (1971) 42-50, 51-54.

\bibitem{Pa07}
D.~Paul, \emph{Asymptotics of large sample covariance matrices of spiked
  population model}, Statistica Sinica \textbf{17} (2007), 1617--1642.

\bibitem{Pe06}
S.~P\'ech\'e, \emph{The largest eigenvalue of small rank perturbations of
  {H}ermitian random matrices}, Prob. Theor. Rel. Fields \textbf{134} (2006),
  127--173.

\bibitem{x} 
K. Rajan and L.F. Abbott, \emph{Eigenvalue spectra of random matrices
for neural networks}, Phys. Rev. Lett. \textbf{97} (2006),
188104.

\bibitem{vii} 
H-J.  Sommers and W. Wieczorek,
\emph{General eigenvalue
correlations for the real Ginibre ensemble}, J. Phys. A \textbf{129},
(2008), 405003(24pp).

\bibitem{St99}
R.P. Stanley, \emph{Enumerative combinatorics}, vol.~2, Cambridge University
  Press, 1999.

\bibitem{xii}
M. Timme, T. Geisel, and F. Wolf, \emph{Speed of synchronization
in complex networks of neural oscillators: analytic results
based on random matrix theory}, Chaos \textbf{16}, (2006) 015108

\bibitem{Wa78}
K.W. Wachter, \emph{The strong limits of random matrix spectra for sample
  matrices of independent elements}, Annal. Prob. \textbf{6} (1978), 1--18.

\bibitem{Wa07}
D.~Wang, \emph{Spiked models in {W}ishart ensembles}, arXiv:0804.0889.

\end{thebibliography}

\providecommand{\bysame}{\leavevmode\hbox to3em{\hrulefill}\thinspace}
\providecommand{\MR}{\relax\ifhmode\unskip\space\fi MR }
\providecommand{\MRhref}[2]{%
  \href{http://www.ams.org/mathscinet-getitem?mr=#1}{#2}
}
\providecommand{\href}[2]{#2}

\end{document}